%% file: main.tex
\documentclass[]{acmart}

\input{sections/_preamble.tex}
\input{sections/_template_args.tex}

\input{sections/_command.tex}

\input{sections/_package.tex}

\usepackage{tikz}
\usetikzlibrary{shapes, arrows, positioning}
\usepackage{orcidlink}

\definecolor{preprocessfill}{HTML}{e1d5e7}
\definecolor{preprocessdraw}{HTML}{9673a6}
\definecolor{featureprocfill}{HTML}{ffe6cc}
\definecolor{featureprocdraw}{HTML}{d79b00}
\definecolor{featureextfill}{HTML}{dae8fc}
\definecolor{featureextdraw}{HTML}{6c8ebf}
\definecolor{featureencfill}{HTML}{d5e8d4}
\definecolor{featureencdraw}{HTML}{82b366}
\definecolor{featurefusfill}{HTML}{f8cecc}
\definecolor{featurefusdraw}{HTML}{b85450}
\definecolor{featuredecfill}{HTML}{d5e8d4}
\definecolor{featuredecdraw}{HTML}{82b366}
\definecolor{renderfill}{HTML}{fff2cc}
\definecolor{renderdraw}{HTML}{d6b656}
\definecolor{modelfill}{HTML}{f5f5f5}

\begin{document}

\title{A conversational gesture synthesis system based on emotions and semantics}

\input{sections/0_authors.tex}

\input{sections/0_abstract.tex}

\input{sections/0_article_info.tex}


\begin{teaserfigure}
	\centering
  \includegraphics[width=\textwidth]{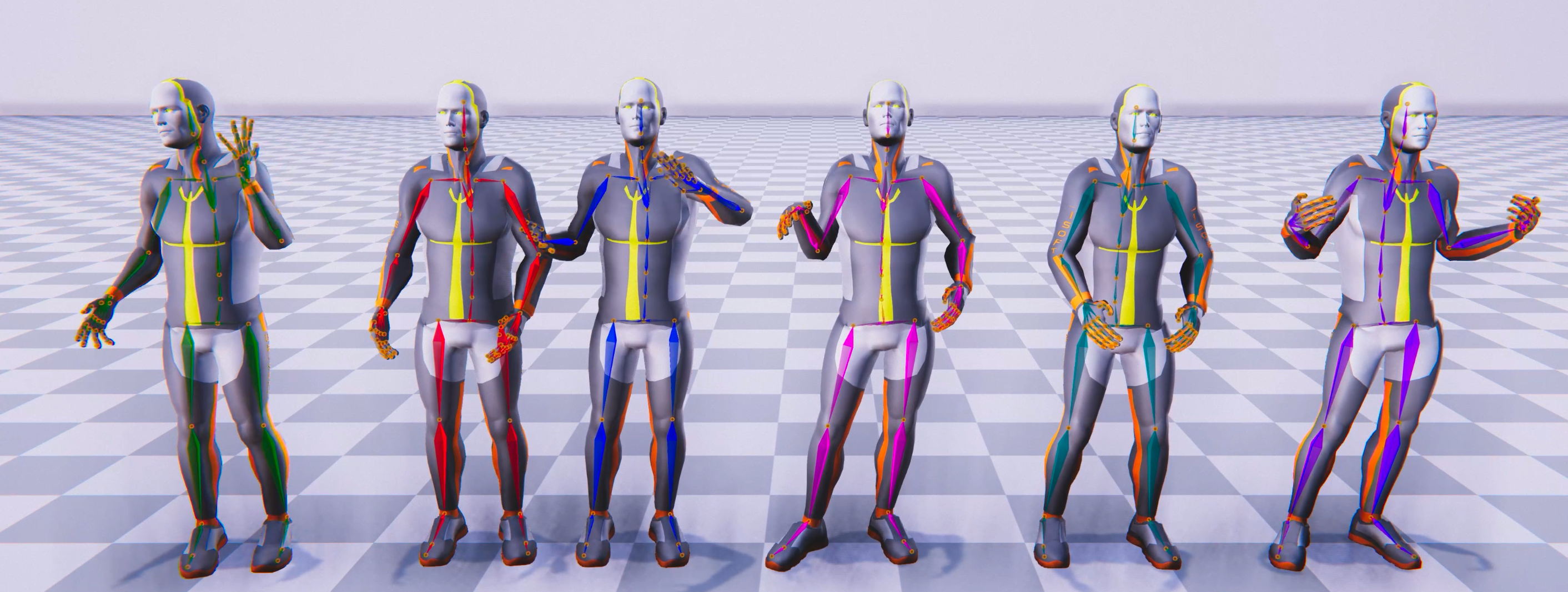}
  \caption{Gesture generation result in various emotion, speech and text}
  \label{fig:teaser}
\end{teaserfigure}

\maketitle

\input{sections/1_introduction}
\input{sections/2_related_work}
\input{sections/3_system_overview}

\input{sections/4_data_preparation}

\input{sections/7_results}
\input{sections/8_conclusion}

\input{sections/x_ack.tex}

\bibliographystyle{ACM-Reference-Format}
\bibliography{reference}

\appendix
\renewcommand{\sectionautorefname}{Appendix}
\renewcommand{\subsectionautorefname}{Appendix}
\renewcommand{\subsubsectionautorefname}{Appendix}
\input{sections/appendix1.tex}
\input{sections/appendix2.tex}

\end{document}

%% file: sections/_preamble.tex
\usepackage{bm}
\usepackage{bbm}

\usepackage{graphicx}
\usepackage{tabularx}
\usepackage{subcaption}
\usepackage{multirow}
\usepackage{amsmath}
\usepackage{mathrsfs}

\usepackage{xspace}

\newcommand{\keep}[1]{}
\newcommand{\old}[1]{}

\usepackage{url}
\usepackage{hyperref}

\usepackage{algorithm}
\usepackage[noend]{algpseudocode}
\makeatletter
\def\BState{\State\hskip-\ALG@thistlm}
\makeatother

\usepackage{fancyhdr}

%% file: sections/_template_args.tex
\AtBeginDocument{%
  \providecommand\BibTeX{{%
    \normalfont B\kern-0.5em{\scshape i\kern-0.25em b}\kern-0.8em\TeX}}}

%% file: sections/_command.tex
\newcommand{\bx}{\mathbf{x}}

\renewcommand{\sectionautorefname}{Section}

\renewcommand{\subsectionautorefname}{Subsection}
\renewcommand{\subsubsectionautorefname}{Subsubsection}

%% file: sections/_package.tex
\usepackage{multicol}
\usepackage{float}

%% file: sections/0_authors.tex
\author{Thanh Hoang-Minh\,\orcidlink{0009-0007-0898-5923}}
\email{hmthanhgm@gmail.com}

\affiliation{
  \institution{OpenHuman.AI}
  \city{Ho Chi Minh City}
  \country{Vietnam}
}

\affiliation{
  \institution{Department of Information Technology, VNUHCM -- University of Science}
  \streetaddress{227 Nguyen Van Cu Street, Cho Quan Ward}
  \city{Ho Chi Minh City}
  \postcode{70000}
  \country{Vietnam}
}

%

\renewcommand{\shortauthors}{Thanh Hoang-Minh}

%% file: sections/0_abstract.tex
\begin{abstract}
	Along with the explosion of large language models, improvements in speech synthesis, advancements in hardware, and the evolution of computer graphics, the current bottleneck in creating digital humans lies in generating character movements that correspond naturally to text or speech inputs.
	
	In this work, we present \textbf{OHGesture}, a diffusion-based gesture synthesis framework for generating expressive co-speech gestures conditioned on multimodal signals - text, speech, emotion, and seed motion. Built upon the DiffuseStyleGesture model, OHGesture introduces novel architectural enhancements that improve semantic alignment and emotional expressiveness in generated gestures. Specifically, we integrate fast text transcriptions as semantic conditioning and implement emotion-guided classifier-free diffusion to support controllable gesture generation across affective states. To visualize results, we implement a full rendering pipeline in Unity based on BVH output from the model. Evaluation on the ZeroEGGS dataset shows that OHGesture produces gestures with improved human-likeness and contextual appropriateness. Our system supports interpolation between emotional states and demonstrates generalization to out-of-distribution speech, including synthetic voices - marking a step forward toward fully multimodal, emotionally aware digital humans.
	\footnote{This manuscript is provided for preview purposes only and has not been published.}
\end{abstract}

%% file: sections/0_article_info.tex
\ccsdesc[500]{Computing methodologies~Animation}
\ccsdesc[300]{Computing methodologies~Natural language processing}
\ccsdesc[300]{Computing methodologies~Neural networks}

%% file: sections/1_introduction.tex
\section{INTRODUCTION}
\label{sec:introduction}

Every day, billions of people around the world look at RGB screens, and the output displayed on these screens is the result of various software systems. Therefore, the rendering of each pixel on the screen and the realistic simulation of images have been a focus of computer graphics scientists since the 1960s, particularly in the simulation of human figures or digital human.

Today, computer graphics technology can realistically simulate many complex objects such as water, roads, bread, and even human bodies and faces with incredible detail, down to individual hair strands, pimples, and eye textures. In 2015, using 3D scanning techniques \cite{metallo2015scanning} to capture all angles of the face and light reflection, researchers were able to recreate President Obama's face on a computer with high precision, making it almost indistinguishable from the real thing.

Artificial intelligence (AI) has shown remarkable results in recent years, not only in research but also in practical applications, such as ChatGPT and Midjourney, showcasing vertical and horizontal growth in various fields. Although computer graphics can construct highly realistic human faces, gesture generation has traditionally relied on Motion Capture from sensors, posing significant challenges in building an AI system that learns from data. Generating realistic beat gestures is challenging because gestural beats and verbal stresses are not strictly synchronized, and it is complicated for end-to-end learning models to capture the complex relationship between speech and gestures.

With the success of large language models in text processing and the advancement of Computer-generated imagery (CGI) in producing nearly indistinguishable human faces, combined with the increasing ease and accuracy of human speech synthesis, gesture generation through AI has become one of the main bottlenecks in developing interactive digital human.

\subsection{Problem Data}
\label{sec:Data}


\begin{figure}[h]
	\centering
	\includegraphics[height=8cm]{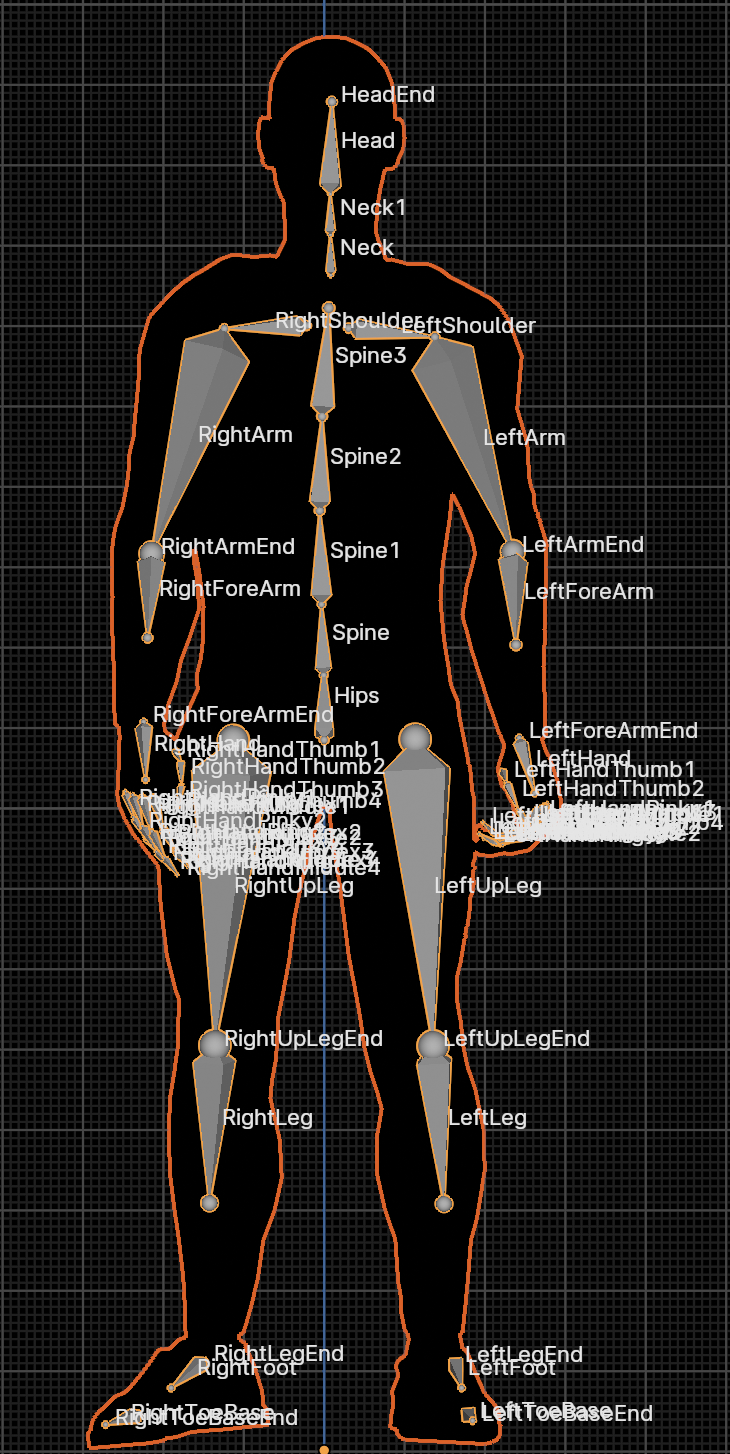}
	\caption{\small Skeleton and joint names of single frame}
	\label{fig:Skeleton}
\end{figure}

\subsubsection{Skeleton Structure of Gestures}

A gesture is defined as the movement of a character's entire body over time, as shown in \autoref{fig:Skeleton}, captured frame by frame. In computer graphics, character motion is represented as bone-specific movements, including hands, legs, head, spine, etc. The full character skeleton structure is presented in Appendix \autoref{appendix:BVHSkeleton}.

Motion data is captured using motion capture systems using cameras and specialized sensors. The output is typically stored in BVH (Biovision Hierarchy) files.

A BVH file consists of two main parts: \texttt{HIERARCHY} and \texttt{MOTION}. The \texttt{HIERARCHY} section is structured as a tree containing the skeleton’s initial positions and names. The \texttt{MOTION} section contains the movement data for the entire skeleton frame-by-frame. Each BVH file includes frame rate (fps) and total frame count. Details are presented in Appendix \autoref{appendix:BVHStructure}.

\subsubsection{Motion Structure of Gestures}

Gesture motion data, as illustrated in \autoref{fig:Skeleton}, or the \texttt{MOTION} section of a BVH file, contains position and rotation information per frame. Each frame is a skeleton of 75 bones: $\{ \mathbf{b}_{1}, \mathbf{b}_{2}, \cdots , \mathbf{b}_{75} \}$, where each bone has position $\{ p_{x}, p_{y}, p_{z} \}$ and rotation $\{ r_{x}, r_{y}, r_{z} \}$.

The output of gesture generation is a sequence of bone rotations per frame. The generated gestures are evaluated based on naturalness, human-likeness, and contextual appropriateness.

The skeleton's position and rotation data are preprocessed into a feature vector $\mathbf{g} \in \mathbb{R}^{D}$ with $D = 1141$. The learning data becomes $\bx \in \mathbb{R}^{M \times D}$. The preprocessing pipeline is detailed in \autoref{appendix:BVHData}.

\subsection{Problem Statement}
\label{sec:ProblemStatement}

The ultimate goal is to produce a sequence of gestures that reflect the motion of the skeleton frame by frame. This can be approached via classification, clustering, or regression. Gesture generation is approached in this work as a regression-based prediction problem, wherein the next sequence is generated conditioned on the current gesture input.

\begin{figure}[h]
	\centering
 	\includegraphics[width=\linewidth]{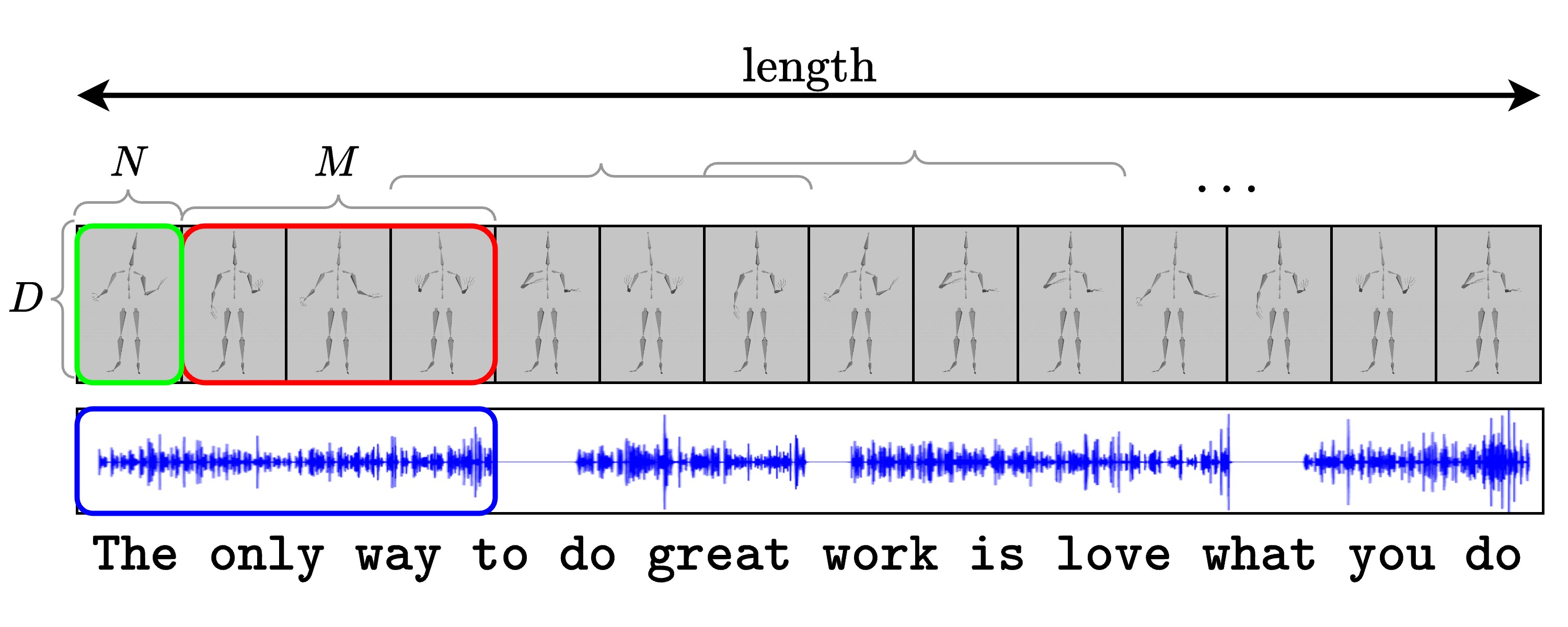}
	\caption{A gesture sequence: the first $N$ frames are used as seed gesture $\mathbf{s}$, and the remaining $M$ frames are to be predicted}
	\label{fig:GestureSeries}
\end{figure}

Each gesture sequence is labeled with an emotion. A key novelty of our approach is pairing the gesture sequence with both the original speech and the corresponding text (transcribed from the speech).

The objective is to build a model that predicts $M$ future frames from the given inputs: seed gesture $\mathbf{s} \in \mathbb{R}^{1:N \times D}$, speech $\mathbf{a}$, text $\mathbf{v}$, and emotion $\mathbf{e}$.

The model prediction is $\hat{\mathbf{x}} \in \mathbb{R}^{1:M \times D}$, which is compared against ground-truth gesture $\mathbf{x} \in \mathbb{R}^{1:M \times D}$.

\begin{multicols}{2}
	
	\textbf{Input}
	
	\begin{itemize}
		\item Seed gesture sequence: $\mathbf{s} \in \mathbb{R}^{1:N \times D}$
		\item Speech signal: $\mathbf{a}$
		\item Text: $\mathbf{v}$
		\item Emotion: $\mathbf{e}$ 
		
		{\small
			(\texttt{Happy},  \texttt{Sad},  \texttt{Neutral}, \texttt{Angry}, \texttt{Old}, \texttt{Funny})
		}
	\end{itemize}
	
	\columnbreak
	
	\textbf{Predicted Output}
	\begin{itemize}
		\item Predicted gesture sequence:
		$\hat{\mathbf{x}} \in \mathbb{R}^{1:M \times D}$
	\end{itemize}
	
	\textbf{Ground Truth}
	\begin{itemize}
		\item Ground truth gesture: $ \mathbf{x}  \in \mathbb{R}^{1:M \times D}$
	\end{itemize}
	
\end{multicols}

\subsection{Challenges}
\label{sec:difficult}

There are several challenges in building a model that can learn human-like conversational gesture patterns:

\begin{enumerate}
	\item \textit{Limited and low-quality data:} Creating large-scale, high-quality datasets for motion capture is extremely costly in the industry.
	
	\item \textit{Inconsistent context between modalities:} Text datasets are more abundant than speech, and speaker attribution is often missing. Synchronization between speech and emotional tone is also lacking. Additionally, training texts span many unrelated topics.
	
	\item \textit{Imbalanced feature distributions:} Current datasets are biased toward English-speaking gestures, with imbalanced gesture distributions between speaking, questioning, and silent states.
	
	\item \textit{High computational cost due to multimodal input:} The model must encode text, speech, and 3D pose data, increasing the computational load during both training and inference. Reducing input information also degrades performance.
	
	\item \textit{Sequential preprocessing steps:} Although human-computer interaction is most effective through speech and keyboard input, processing the text and speech input for gesture generation must be done sequentially. In real-world applications, inference latency is critical, and users cannot wait long. Rendering the gestures on screen must also be optimized for speed.
\end{enumerate}

\large \textbf{Contributions}

 In summary, our main contributions in this paper are:
 
\begin{itemize}
	\item From existing datasets, speech is transcribed into text and used as additional semantic features for training.
	
	\item Based on the DiffuseStyleGesture model, we extend the conditional denoising process to include text features.
	
	\item In this work, we use Unity for rendering, data extraction, and visualization of gesture generation results.
	
	\item A rendering pipeline is designed and implemented in this paper, with the system demonstrated using Unity.
\end{itemize}


%% file: sections/2_related_work.tex
\section{RELATED WORK}
\label{sec:related_work}

Gesture generation, like other machine learning tasks, has been explored using both traditional rule-based methods and modern data-driven approaches. We first examine the fundamental characteristics of gestures (\autoref{sec:relationspeechandgesture}) as a basis for modeling their relationship with text and speech. \autoref{fig:CommonStage} illustrates the common processing stages shared across various gesture generation techniques.

\subsection{Gesture Characteristics}
\label{sec:relationspeechandgesture}

According to linguistics, gestures can be categorized into six main groups: adaptors, emblems, deictics, iconics, metaphorics, and beats \cite{ekman1969repertoire}, \cite{sebeok2011advances}. Among them, beat gestures do not carry direct semantic meaning but play an important role in synchronizing rhythm between speech and gesture \cite{kipp2005gesture}, \cite{sebeok2011advances}. However, the rhythm between speech and beat gestures is not fully synchronized, making the temporal alignment between them complex to model \cite{mcclave1994gestural}, \cite{bhattacharya2021speech2affectivegestures}, \cite{kucherenko2020gesticulator}, \cite{yoon2020speech}.

Gestures interact with various levels of information in speech \cite{sebeok2011advances}. For instance, emblematic gestures like a thumbs-up are usually linked to high-level semantic content (e.g., “good” or “excellent”), while beat gestures often accompany low-level prosodic features such as emphasis. Previous studies typically extract features from the final layer of the speech encoder to synthesize gestures \cite{alexanderson2020style}, \cite{bhattacharya2021speech2affectivegestures}, \cite{kucherenko2021large}, \cite{qian2021speech}, \cite{yoon2022genea}. However, this approach may blend information from different levels, making it hard to disentangle rhythm and semantics.

As shown in linguistic studies \cite{kipp2005gesture}, \cite{neff2008gesture}, \cite{webb1997linguistic}, gestures in daily communication can be decomposed into a limited number of semantic units with various motion variations. Based on this assumption, speech features are divided into two types: high-level features representing semantic units and low-level features capturing motion variations. Their relationships are learned across different layers of the speech encoder. Experiments demonstrate that this mechanism can effectively disentangle features at different levels and generate gestures that are both semantically and stylistically appropriate.

\subsection{Overview of Gesture Generation Methods}
\label{sec:relatedwork}

\subsubsection{Rule-Based Methods}

These methods rely on clearly defined rules, which are manually crafted to determine how the system processes inputs to produce outputs.

\textbf{Methods}: Representative rule-based methods include the \textit{Robot behavior toolkit} \cite{huang2012robot} and \textit{Animated conversation} \cite{cassell1994animated}. These approaches typically map speech to gesture units using handcrafted rules. Rule-based systems allow for straightforward control over model outputs and provide good interpretability.  
However, the cost of manually designing these rules is prohibitive for complex applications requiring the processing of large-scale data.

\subsubsection{Statistical Methods}

These methods rely on data analysis, learning patterns from datasets, and using probabilistic models or mathematical functions for prediction. The approach involves optimizing model parameters to fit the data.

\textbf{Methods}: Like rule-based methods, data-driven methods also map speech features to corresponding gestures. However, instead of manual rules, they employ automatic learning based on statistical data analysis.

Representative statistical approaches include \textit{Gesture controllers} \cite{levine2010gesture}, \textit{Statistics-based} \cite{yang2020statistics}, which use probabilistic distributions to find similarities between speech and gesture features. \textit{Gesture modeling} \cite{neff2008gesture} constructs probabilistic models to learn individual speaker styles.

\subsubsection{Deep Learning Methods}

These methods utilize multi-layer perceptrons (MLPs) to automatically extract features from raw data and learn complex data representations through parameter optimization.


Deep learning-based gesture generation methods can be divided into two main groups: likelihood-based models and implicit generative models \cite{song2021score}.

\textbf{Likelihood-Based Models}

These models work by maximizing the likelihood of observed data given model parameters $\theta$. The objective is to find the optimal parameters $\theta'$ by modeling the probability $p(\mathbf{x})$ of the data, where $\mathbf{x}$ represents the gesture sequence.

\textbf{Methods}: The application of deep learning to gesture generation has evolved alongside the development of deep learning models, including RNNs, LSTMs, and Transformers. Representative likelihood-based methods include:

\begin{itemize}
	\item \textit{Gesticulator} \cite{kucherenko2020gesticulator}, which uses a Multilayer Perceptron (MLP) to encode text and audio features, with BERT-derived vectors used as text features.
	
	\item \textit{HA2G} \cite{liu2022learning} builds a hierarchical Transformer-based model to learn multi-level correlations between speech and gestures, from local to global features.
	
	\item \textit{Gesture Generation from Trimodal Context} \cite{yoon2020speech} uses an RNN architecture and treats gesture generation as a translation task in natural language processing.
	
	\item \textit{DNN} \cite{chiu2015predicting} combines LSTM and GRU to build a classifier neural network that selects appropriate gesture units based on speech input.
	
	\item \textit{Cascaded Motion Network (CaMN)} \cite{liu2022beat} introduces the BEAT dataset and a waterfall-like model. Speaker information, emotion labels, text, speech, and gesture features are processed through layers to extract latent vectors. In the fusion stage, CaMN combines features sequentially: speaker and emotion features are merged first, followed by integration with latent vectors of text, speech, and gestures.
	
	\item \textbf{Motion Graph}: In \textit{Gesturemaster} \cite{zhou2022gesturemaster}, a semantic graph is constructed where words in a sentence are connected based on semantic relationships. The model then selects the most relevant nodes and edges in the graph to represent gestures.
\end{itemize}


\vfill
\begin{figure*}[htbp]
	\centering
	\includegraphics[width=0.9\linewidth]{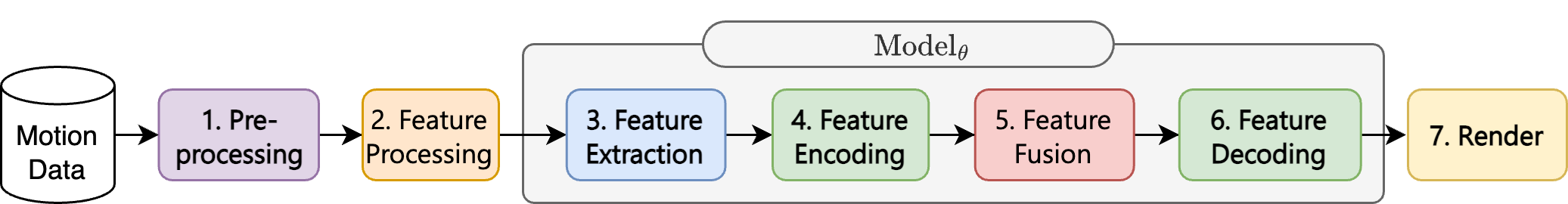}
	\caption{Common stages in gesture generation models.}
	\label{fig:CommonStage}
\end{figure*}

\textbf{Implicit Generative Models}
\label{sec:ImplicitGenerativeModels}

Implicit generative models learn the data distribution without explicitly modeling the probability density function $p(\bx)$. Instead of directly computing $p(\bx)$, the model learns a mapping $G_{\theta}: \mathcal{Z} \to \mathcal{X}$ by matching the distributions between real and generated data $\mathbf{x} = G_\theta(\mathbf{z}), \quad \mathbf{z} \sim p_z(\mathbf{z})$. Here, $\mathcal{Z}$ denotes the input noise space, typically with a simple distribution \(p_{z}\) (e.g., Gaussian, Uniform), and \(\mathcal{X}\) is the space of real data, which in this case is gesture sequences \(\mathbf{x}\).

In gesture generation, to incorporate conditions such as speech or text labels, implicit generative models often introduce a condition $\mathbf{c}$ representing the context of the task, resulting in the conditional generation: $\mathbf{x} = G_\theta(\mathbf{z}, \mathbf{c})$. This context may include speech, text, speaker ID, style, emotion, or initial gestures.

A typical example is the generative adversarial network (GAN) and Diffusion model, where data is synthesized by transforming an initial simple distribution (e.g., Gaussian) into the target data distribution.

\textbf{Methods}: Representative implicit generative models include:

\begin{itemize}
	\item \textit{MoGlow} \cite{henter2020moglow} uses Normalizing Flows to maintain motion consistency and compatibility, while providing control via input parameters. This allows users to generate new motions or modify existing ones by adjusting control parameters.
	
	\item \textbf{GAN}: \textit{GRU-based WGAN} \cite{wu2021probabilistic} utilizes Wasserstein GAN to evaluate and improve the quality of gesture synthesis. The model focuses on optimizing the Wasserstein loss, mitigating mode collapse typically seen in traditional GANs. GRUs process speech data and convert it into usable gesture features, which are then fed into the WGAN for evaluation and refinement.
	
	\item \textbf{VAE}: \textit{FreeMo} \cite{xu2022freeform} employs a VAE to decompose gestures into pose modes and rhythmic motions. Gesture poses are randomly sampled using conditional sampling in the VAE's latent space, while rhythmic motions are generated from...
	
	\item \textbf{VQ-VAE}: \textit{Rhythmic Gesticulator} \cite{ao2022rhythmic} preprocesses speech segments based on beats, dividing speech into smaller parts and representing them as blocks with normalized dimensions. Similar to the VQ-VAE approach, normalized gesture sequences are quantized into discrete gesture lexicons. The model learns the gesture vocabulary conditioned on the previous gesture, gesture style, and speech. It then reconstructs the gesture sequence via denormalization. Unlike generative models like GANs or Diffusion, VQ-VAE focuses more on compression rather than direct generation.
	
	\item \textbf{Diffusion}: Diffusion models focus on generating new data from noisy inputs by progressively denoising toward the original data. Diffusion-based approaches will be presented in \autoref{sec:diffusionbase}.
\end{itemize}

\subsection{Common Stages in Deep Learning Approaches for Gesture Generation}
\label{sec:commonstage}

As presented in \autoref{sec:Data}, gestures consist of sequences of 3D point coordinates. For each dataset, the number of bones per frame may vary.

Deep learning approaches to gesture generation are implemented using various techniques. However, we generalize the process into the following main stages, illustrated in \autoref{fig:CommonStage}:

\begin{enumerate}
	\item \textbf{Preprocessing}: In the preprocessing stage, data such as speech segments, gesture sequences, and text are read and digitized into vectors or matrices that represent raw data information. Depending on the specific learning method, the selected initial data features may vary.
	
	\item \textbf{Feature Processing}: In this stage, raw data such as speech and text are embedded into feature vectors. Different methods use different embedding models. The way gesture sequences are represented as feature vectors also varies across methods.
	
	\item \textbf{Feature Extraction}: This stage uses linear transformation layers or CNN layers to extract features from the data. Text and speech features, after being processed, may be further passed through feature extraction layers to generate representation vectors corresponding to the input modalities.
	
	\item \textbf{Feature Encoding}: In this stage, gesture, emotion, and speech vectors are encoded into a lower-dimensional latent space to facilitate learning the correlations among modalities in the feature fusion stage.
	
	\item \textbf{Feature Fusion}: In this stage, features from speech, text, gestures, and other information are combined, typically using concatenation, fully connected layers, or operations such as vector addition or subtraction in the latent space.
	
	\item \textbf{Feature Decoding}: In this stage, latent vectors are decoded or upsampled back to their original dimensionality.
	
	\item \textbf{Rendering}: Once the output vectors are restored to their original size, they are converted back into BVH files and rendered using software such as Blender or Unity to visualize character motion.
\end{enumerate}

\subsection{Diffusion-based Model for Gesture Generation}
\label{sec:diffusionbase}

\begin{itemize}
	\item \textit{MotionDiffuse} \cite{zhang2022motiondiffuse} employs a conditional Diffusion model, with conditions based solely on text, excluding audio. Additionally, the model predicts noise rather than directly predicting the original gesture sequence. MotionDiffuse utilizes Self-Attention and Cross-Attention layers to model the correlation between textual features and gesture features during \textit{Stage 5. Feature Fusion} (\autoref{fig:CommonStage}).
	
	\item \textit{Flame} \cite{kim2023flame} applies a Diffusion model with a Transformer-based architecture. In \textit{Stage 2. Feature Processing} (\autoref{fig:CommonStage}), it uses the pre-trained RoBERTa model to embed the text into textual feature vectors, which serve as the conditioning input. During \textit{Stage 5. Feature Fusion} (\autoref{fig:CommonStage}), the text is used as the $\texttt{CLS}$ token prepended to the gesture sequence before passing through the Transformer Decoder. Similar to other methods, the model predicts the added noise rather than the original gesture sequence.
	
	\item \textit{DiffWave} \cite{kong2020diffwave} is a noise-predicting Diffusion model in which the time steps pass through multiple Fully Connected layers and a Swish activation function before feature fusion. It uses a dilated convolutional architecture inherited from WaveNet. DiffWave enables better representation of speech, improving the effectiveness of conditioning for the Diffusion model.
	
	\item \textit{Listen, Denoise, Action} \cite{alexanderson2022listen} builds upon DiffWave \cite{kong2020diffwave}, replacing the dilated convolution layers with a Transformer, and integrating Conformer modules to enhance model performance.
	
	\item \textit{DiffSHEG} \cite{chen2024diffsheg} employs a Diffusion model; in \textit{Stage 2. Feature Processing}, it uses HuBERT to encode the audio signal. The model treats facial expressions as a signal for gesture generation and achieves real-time fusion of both facial expressions and gestures.
	
	\item \textit{GestureDiffuCLIP} \cite{ao2023gesturediffuclip} uses a Diffusion model conditioned on text, leveraging Contrastive Learning with CLIP to integrate text features and control gesture styles. Similar to other prompt-based approaches such as StableDiffusion or Midjourney, it treats text as prompts to learn gestures from descriptive sentences.
	
	\item \textit{Freetalker} \cite{yang2024freetalker} trains a Diffusion model on multiple datasets to generate speaker-specific gestures conditioned on speech and text. Unlike Transformer-based methods, Freetalker employs an Attention-based Network to model the correlation between textual, auditory, and gesture features during \textit{Stage 5. Feature Fusion} (\autoref{fig:CommonStage}).
\end{itemize}

\begin{figure}[htbp]
	\centering
	\includegraphics[width=\linewidth]{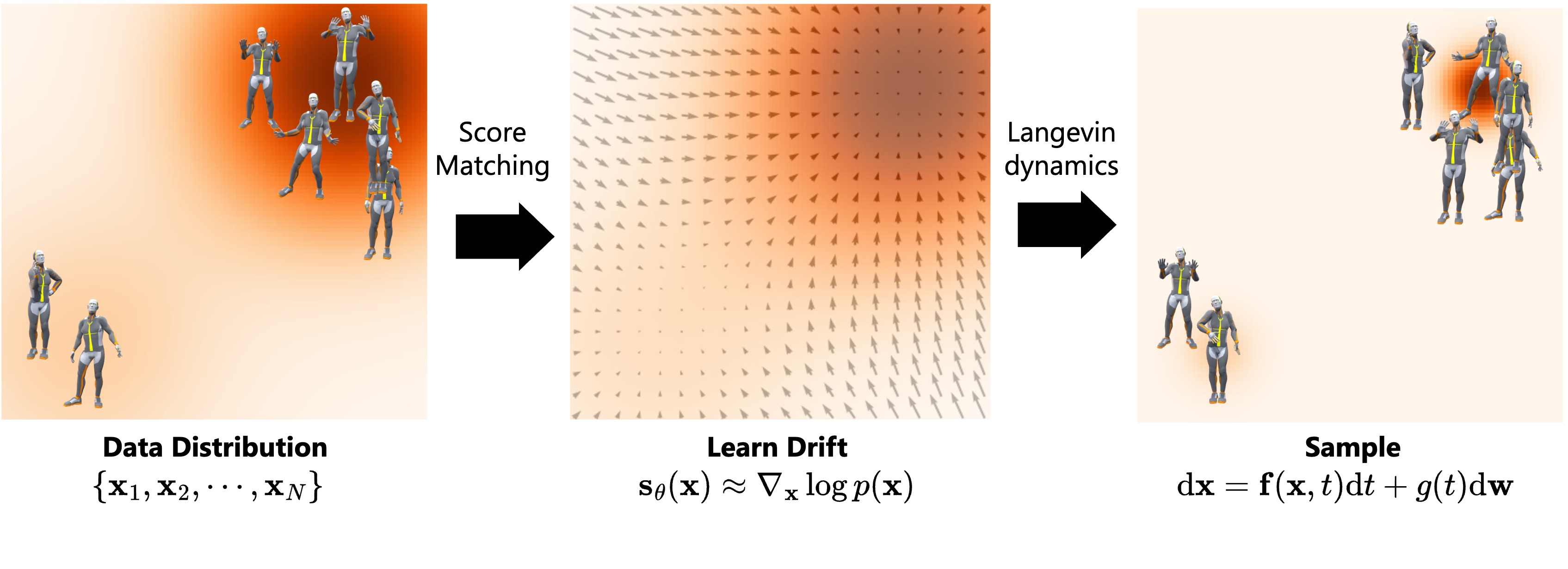}
	\caption{Illustration of the diffusion drift term in gesture generation. The figure demonstrates how the learned drift guides the reverse diffusion process to synthesize temporally coherent and semantically relevant gestures from noise.}
	\label{fig:ScoreMatching}
\end{figure}

\vspace{-0.2cm}
\subsubsection{Selected Diffusion-based Methods}

\begin{itemize}
	\item \textit{MDM} \cite{tevet2022human} applies conditional Diffusion to gesture generation, using CLIP (Contrastive Language–Image Pre-training) embeddings of descriptive text as conditions. MDM adopts a Transformer-based architecture to reconstruct original gesture data. Like other text-based Diffusion approaches, in \textit{Stage 3. Feature Extraction} (\autoref{fig:CommonStage}), the input text is randomly masked to hide certain segments, enabling the model to learn the importance of each part for various gestures. 
	
	In \textit{Stage 5. Feature Fusion} (\autoref{fig:CommonStage}), the text is prepended as a $\texttt{[CLS]}$ token to the gesture sequence before passing through the Transformer Encoder, where self-attention models the relationship between the text and each gesture frame. MDM predicts the original gesture data rather than noise.
	
	\item \textbf{DiffuseStyleGesture} \cite{yang2023diffusestylegesture} extends \textit{MDM} \cite{tevet2022human} by incorporating audio, initial gestures, and style as conditioning inputs. In \textit{Stage 1. Preprocessing} (\autoref{fig:CommonStage}), the model processes coordinate vectors to obtain feature vectors of dimension $D=1141$ per frame. In \textit{Stage 2. Feature Processing} (\autoref{fig:CommonStage}), DiffuseStyleGesture uses WavLM for audio embedding. In \textit{Stage 5. Feature Fusion} (\autoref{fig:CommonStage}), it improves upon MDM by applying Cross-Local Attention prior to the Transformer Encoder.
\end{itemize}

%% file: sections/3_system_overview.tex
\section{PROPOSED METHOD}
\label{sec:system_overview}

\begin{figure}[h]
	\centering
	\includegraphics[width=\linewidth]{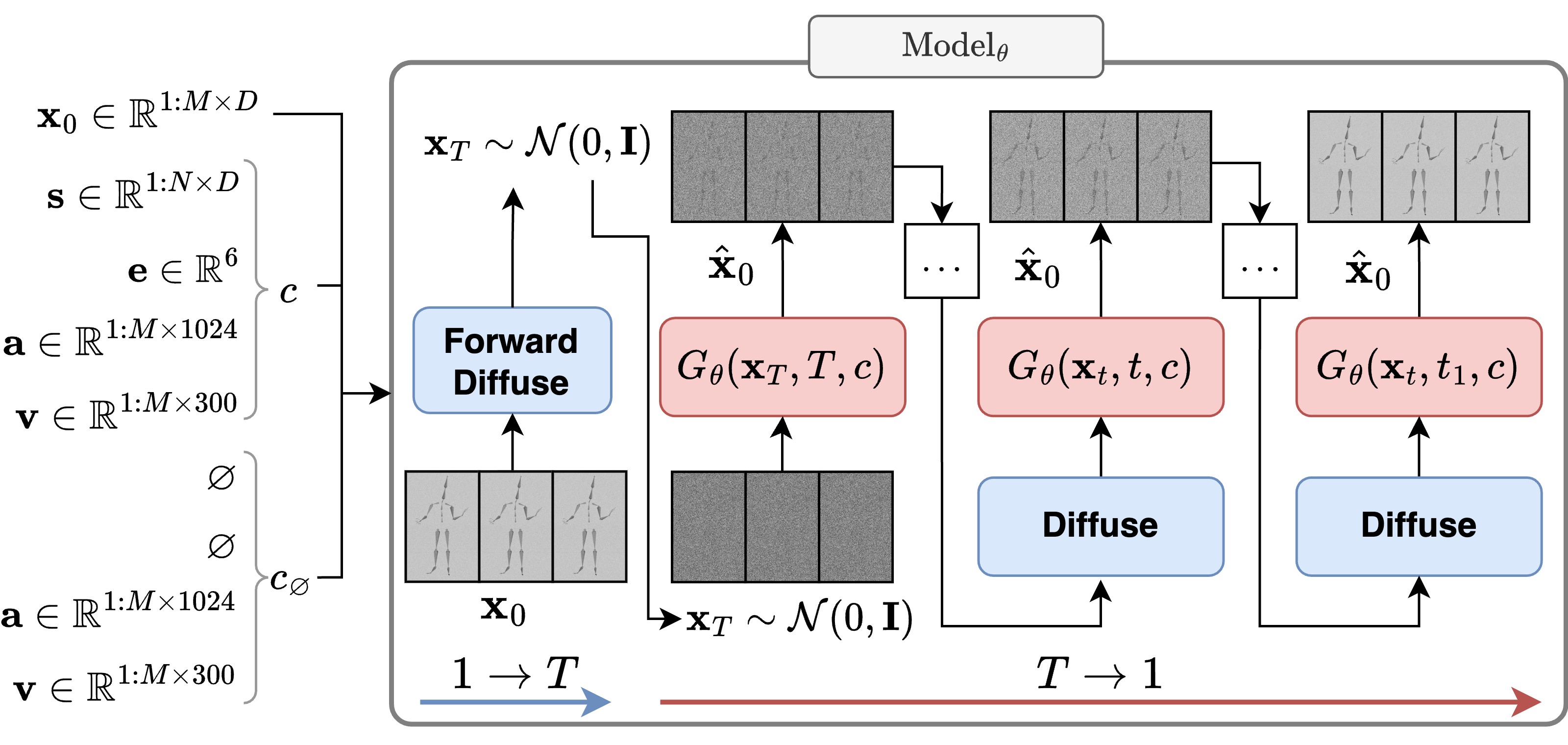}
	\caption{Overview of the OHGesture model}
	\label{fig:TrainingAndSampling}
\end{figure}

The proposed \textbf{OHGesture} model is based on the \textbf{DiffuseStyleGesture} model \cite{yang2023diffusestylegesture}, which applies the Diffusion model \cite{ho2020denoising} with conditional guidance \cite{ho2022classifier} (Classifier-Free Diffusion Guidance) to control features during the denoising process.

The similarities and differences of applying the diffusion model to the gesture generation task compared to image generation are as follows:

\textbf{Similarities}
\begin{itemize}
	\item Uses the Diffusion model on gesture data $\bx^{1:M \times D}$, with $M$ temporal frames and $D=1141$ representing motion coordinates per frame (analogous to image width and height).
	\item Uses conditional Diffusion with $\bx_0$ objective.
	\item In stages \textit{4. Feature Encoding} and \textit{6. Feature Decoding} in \autoref{fig:CommonStage}, the model uses a latent vector of dimension $256$.
\end{itemize}

\textbf{Differences}
\begin{itemize}
	\item Conditional gesture generation:
	\begin{itemize}
		\item Emotional condition: $c = \big[ \mathbf{s}, \mathbf{e}, \mathbf{a}, \mathbf{v} \big]$ and $c_{\varnothing} = \big[ \varnothing, \varnothing, \mathbf{a}, \mathbf{v}\big]$.
		\item Emotion state interpolation between $\mathbf{e}_1, \mathbf{e}_2$ using: $c = \big[ \mathbf{s}, \mathbf{e}_1, \mathbf{a}, \mathbf{v} \big]$ and $c_{\varnothing} = \big[ \mathbf{s}, \mathbf{e}_2, \mathbf{a}, \mathbf{v} \big]$.
	\end{itemize}
	\item In stage \textit{5. Feature Fusion} \autoref{fig:CommonStage}, the model uses Self-Attention: learning the relationship between emotions, seed gestures, and each frame (similar to DALL-E 2's text-image alignment).
	\item In stage \textit{5. Feature Fusion} \autoref{fig:CommonStage}, the model concatenates speech and text (analogous to ControlNet's pixel-wise condition).
\end{itemize}

Here, $\bx_0$ is a sequence of $M$ gesture frames $\mathbf{x} \in \mathbb{R}^{1:M \times D}$ ($D = 1141$), with condition $c = [\mathbf{s}, \mathbf{e}, \mathbf{a}, \mathbf{v}]$ including seed gesture $\mathbf{s}$, emotion $\mathbf{e}$, speech $\mathbf{a}$ corresponding to the gesture, and text $\mathbf{v}$.

The model's objective is to learn parameters $\theta$ of the generative function $G_{\theta}$ with inputs being the noisy gesture matrix $\bx_t \in \mathbb{R}^{1:M \times D}$, timestep $t$, and condition $c$. An overview of the proposed \textbf{OHGesture} model is illustrated in \autoref{fig:TrainingAndSampling}. As with standard diffusion models, it includes two processes: the diffusion process $q$ and the denoising process $p_{\theta}$ with weights $\theta$. The \textit{1. Preprocessing} stage will be presented in \autoref{sec:Preprocessing}.

\subsection{Feature Processing Stage}

In the \textit{Stage 2: Feature Processing} step (\autoref{fig:CommonStage}), the goal is to convert raw input data into matrices or vectors suitable for input into the model.

\begin{itemize}
	\item \textbf{Text} $\mathbf{v} \in \mathbb{R}^{1:M \times 300}$:  
	As discussed in \autoref{sec:diffusionbase}, methods such as \textit{MDM} \cite{tevet2022human} and \textit{DiffuseStyleGesture+} \cite{yang2022DiffuseStyleGestureplus} use gesture-descriptive prompts, similar to those used in Midjourney, as input for the model. However, such prompts are mainly used to cluster gestures rather than align them semantically. 
	
	In contrast, we treat text as a semantic feature that aligns specific segments of text with corresponding gesture segments for digital human generation. Therefore, the contribution in this stage is to use speech data that has already been preprocessed (see \autoref{sec:Preprocessing}) to obtain transcribed text. We use the FastText model \cite{bojanowski2017enriching} to embed this text into vectors, which are then temporally aligned with the number of gesture frames, resulting in a text matrix $\mathbf{v} \in \mathbb{R}^{1:M \times 300}$. For segments without text, zero vectors are used; for segments with vocabulary, the corresponding embedding is assigned for each gesture frame.
	
	\item \textbf{Speech} $\mathbf{a} \in \mathbb{R}^{1:M \times 1024}$:  
	All speech data in `.wav` format is downsampled to $16~\mathrm{kHz}$. The speech corresponding to the gesture segment (4 seconds) is extracted as a waveform vector $\mathbf{a} \in \mathbb{R}^{64000}$. Following DiffuseStyleGesture, this work uses the pre-trained WavLM Large model \cite{Chen_2022} to embed the raw waveform into a high-dimensional vector representing acoustic features. Linear interpolation is then applied to temporally align the latent features from WavLM at $20~\text{fps}$, resulting in the speech matrix $\mathbf{a} \in \mathbb{R}^{1:M \times 1024}$.
	
	\item \textbf{Emotion} $\mathbf{e} \in \mathbb{R}^{6}$:  
	Emotion is represented by one of six classes: $\texttt{Happy}$, $\texttt{Sad}$, $\texttt{Neutral}$, $\texttt{Old}$, $\texttt{Relaxed}$, and $\texttt{Angry}$.  
	Each label is encoded using one-hot encoding to form the vector $\mathbf{e} \in \mathbb{R}^{6}$.
	
	\item \textbf{Seed Gesture} $\mathbf{s} \in \mathbb{R}^{1:N \times D}$:  
	This is the initial gesture sequence composed of $N=8$ frames, with each frame containing joint data for 75 joints. These are processed according to the formula in \autoref{eq:gesturevector} to yield a $D=1141$-dimensional vector.
	
	\item \textbf{Ground Truth Gesture} $\mathbf{x}_{0} \in \mathbb{R}^{1:M \times D}$:  
	This is the ground-truth gesture sequence of $M = 80$ frames (corresponding to 4 seconds at $20~\text{fps}$), which is preprocessed to form the matrix $\mathbf{x}_{0} \in \mathbb{R}^{1:M \times D}$.
\end{itemize}

\subsection{Feature Extraction Stage}
\label{subsec:feature_extraction}

In the \textit{Stage 3: Feature Extraction} step (\autoref{fig:CommonStage}), the goal is to convert the input matrices into latent vectors that capture the semantic content of each modality. This is done by passing the feature data through linear transformation layers or Multilayer Perceptrons (MLPs).

\begin{figure*}[t]
	\centering
	\includegraphics[width=\linewidth]{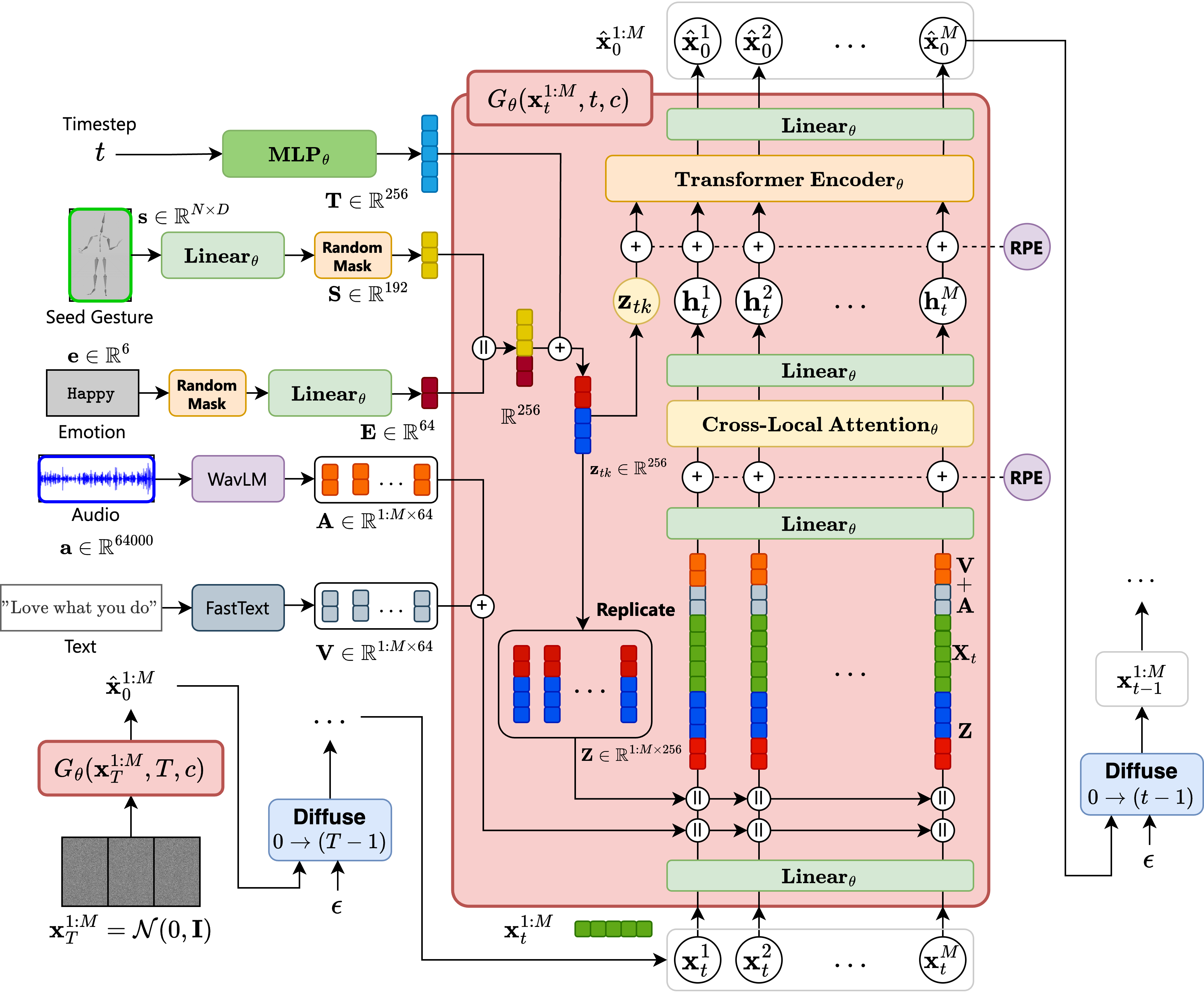}
	\caption{The Model Architecture in OHGesture}
	\label{fig:OHGesture}
\end{figure*}

\begin{itemize}
	\item \textbf{Timestep $\mathbf{T} \in \mathbb{R}^{256}$}: The timestep at each process step $t \in [0, T]$ enables the model to generalize the denoising process across different time steps. The objective is for the model to learn how the values should change with respect to $t$ in order to accurately predict $\bx_0$. The timestep $t$ is initialized using sinusoidal positional encoding: $\text{PE}(t) = \left[ \sin{\left(\frac{t}{10000^{2i / d}}\right)}, \cos{\left(\frac{t}{10000^{2i / d}}\right)} \right]$, and then passed through a Multilayer Perceptron (MLP) to obtain the vector $\mathbf{T} \in \mathbb{R}^{256}$.
	
	\item \textbf{Speech $\mathbf{A} \in \mathbb{R}^{1:M \times 64}$}: The speech feature matrix $\mathbf{a} \in \mathbb{R}^{1:M \times 1024}$ is passed through a linear layer to reduce its dimensionality to a 64-dimensional feature vector, resulting in the matrix $\mathbf{A} \in \mathbb{R}^{1:M \times 64}$.
	
	\item \textbf{Text $\mathbf{V} \in \mathbb{R}^{1:M \times 64}$}: After preprocessing described in \autoref{sec:Preprocessing}, the text is aligned to match the number of frames $M$, resulting in the matrix $\mathbf{v} \in \mathbb{R}^{1:M \times 300}$. It is then passed through a linear transformation to reduce feature dimensionality, yielding the final matrix $\mathbf{V} \in \mathbb{R}^{1:M \times 64}$, aligned with the speech matrix.
	
	\item \textbf{Seed Gesture $\mathbf{S} \in \mathbb{R}^{192}$}: The seed gesture $\mathbf{s} \in \mathbb{R}^{1:N \times D}$ is processed through a linear transformation layer to obtain the vector $\mathbf{S} \in \mathbb{R}^{192}$. During training, $\mathbf{S}$ is also passed through a random masking layer to randomly mask segments of gesture frames in $N$ frames. This allows the model to learn how missing frames impact the final predicted gestures sequence.
	
	\item \textbf{Emotion $\mathbf{E} \in \mathbb{R}^{64}$}: The emotion vector $\mathbf{e} \in \mathbb{R}^{6}$ is passed through a linear transformation layer to produce the feature vector $\mathbf{E} \in \mathbb{R}^{64}$. This is designed to be concatenated with the seed gesture vector $\mathbf{S} \in \mathbb{R}^{192}$ to form a 256-dimensional latent vector.
	
	\item \textbf{Noisy Gesture $\mathbf{x}_{T} \in \mathbb{R}^{1:M \times D}$}: During training, $\mathbf{x}_t$ is the noisy gesture that shares the same dimensionality as the original gesture $\mathbf{x}_0$ and is sampled from a standard normal distribution $\mathcal{N}(0, \mathbf{I})$. The initial noisy gesture $\mathbf{x}_T$ is drawn from a Gaussian distribution, and the subsequent $\mathbf{x}_t, \, t < T$ are produced through iterative noising steps, as illustrated in \autoref{fig:TrainingAndSampling}.
\end{itemize}

\subsection{Feature Encoding Stage}

In the \textit{Stage 4: Feature Encoding} step (\autoref{fig:CommonStage}), the objective is to reduce the dimensionality of the input data to a lower latent space, in order to alleviate computational overhead and avoid the explosion of processing complexity.

The primary data used in the diffusion process is the gesture sequence $\bx_t \in \mathbb{R}^{1:M \times D}$. As illustrated in \autoref{fig:OHGesture}, the gesture sequence with size $M \times D$ is passed through a linear transformation layer $\operatorname{Linear}_{\theta}$ to produce the matrix $\mathbf{X} \in \mathbb{R}^{1:M \times 256}$. This dimensionality reduction of $\bx$ is performed prior to passing the gesture sequence through the Cross-Local Attention and Transformer Encoder layers, in order to compute correlations across multiple modalities.

\subsection{Feature Fusion Stage}

In the \textit{Stage 5: Feature Fusion} step (\autoref{fig:CommonStage}), the goal is to compute inter-feature correlations using concatenation, addition, or attention-based mechanisms.

First, the seed gesture vector $\mathbf{S} \in \mathbb{R}^{192}$ and the emotion vector $\mathbf{E} \in \mathbb{R}^{64}$ are concatenated to form a single vector of size $256$, since $256$ is the chosen hidden dimensionality for computing feature correlations. This vector is then added to the timestep vector $\mathbf{T}$ to form the final vector $\mathbf{z}_{tk} \in \mathbb{R}^{256}$.

\begin{equation}
	\label{eq:ConditionConcat}
	\mathbf{z}_{tk} = \operatorname{concat }(\mathbf{E}\ || \  \mathbf{S}) + \mathbf{T}
\end{equation}

\begin{figure}[h]
	\centering
	\begin{subfigure}{0.42\linewidth}
		\centering
		\includegraphics[width=0.95\linewidth]{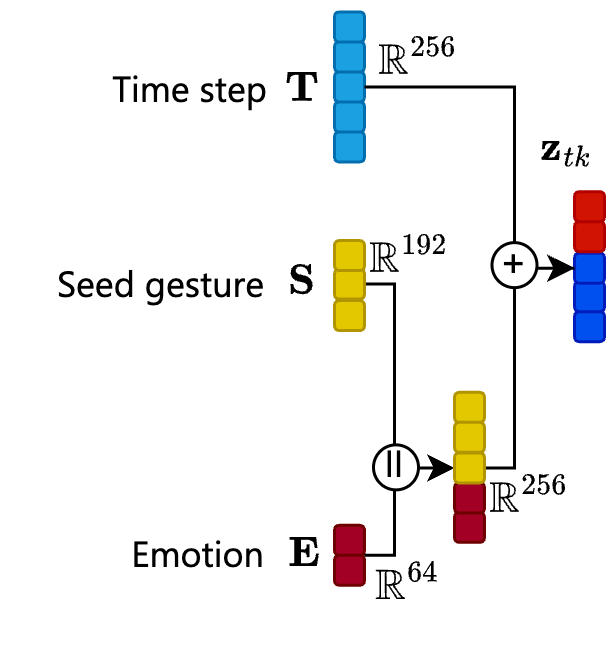}
		\caption{Feature concatenation $\mathbf{z}_{tk}$}
		\label{fig:FeatureFusion}
	\end{subfigure}
	\hfill
	\begin{subfigure}{0.55\linewidth}
		\centering
		\includegraphics[width=0.85\textwidth]{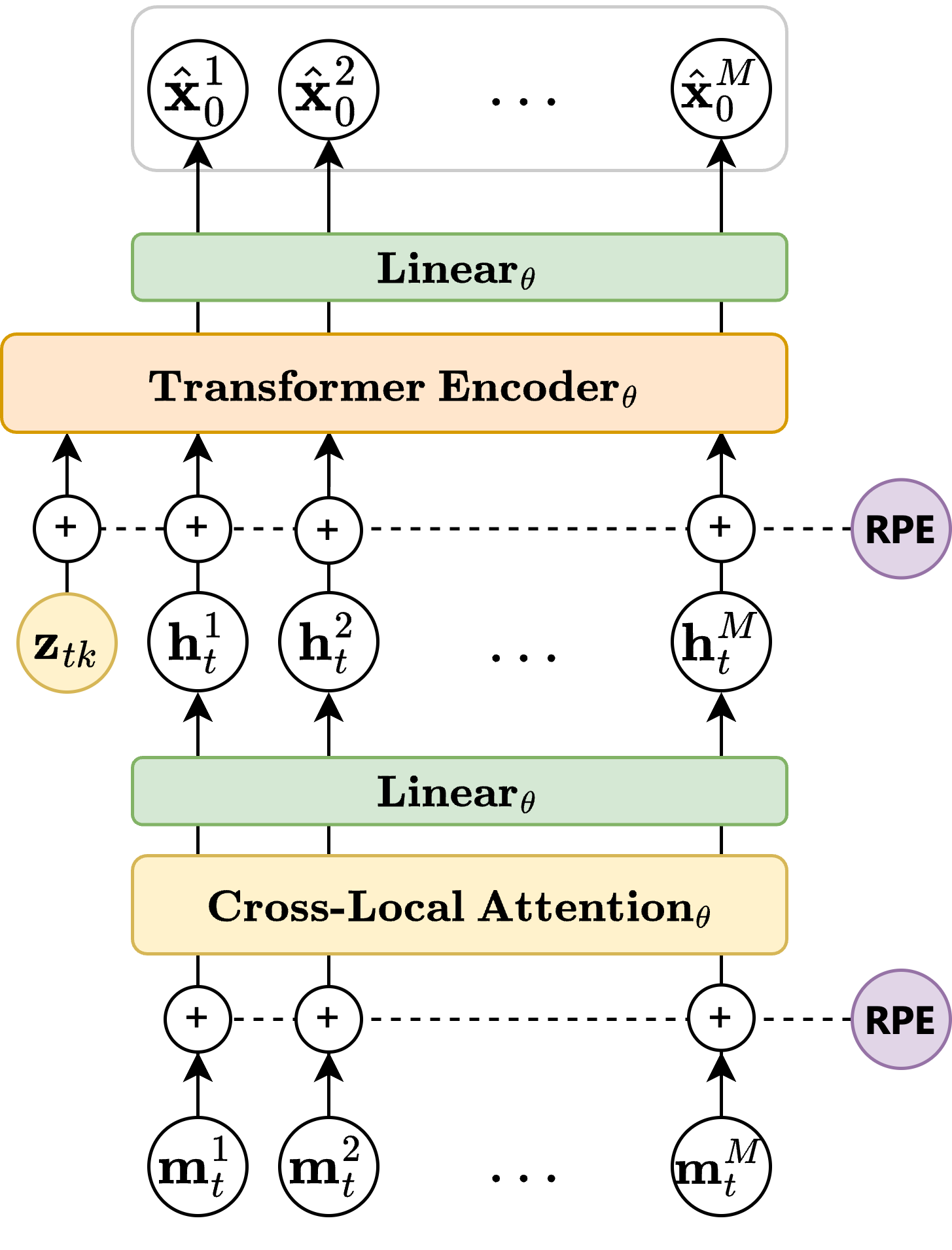}
		\caption{Frame-wise feature-fusion process}
		\label{fig:ZToken}
	\end{subfigure}
\end{figure}

The feature fusion process of $\mathbf{z}_{tk}$ is illustrated in \autoref{fig:FeatureFusion}.

\subsubsection{Frame-wise Feature Integration}

Next, $\mathbf{z}_{tk} \xrightarrow{\operatorname{replicate}} \mathbf{Z}$ is replicated $M$ times to match the dimensionality of $M$ frames, resulting in the matrix $\mathbf{Z} \in \mathbb{R}^{1:M \times 256}$, as shown in \autoref{fig:OHGesture}.

The speech feature matrix $\mathbf{A}$ and the text feature matrix $\mathbf{V}$ are then added together to produce a matrix that combines both modalities. This combined matrix is then concatenated with the gesture feature matrix $\mathbf{X}$. Finally, the result is concatenated with matrix $\mathbf{Z}$ to obtain the full feature matrix $\mathbf{M}$.

\begin{equation}
	\label{eq:FrameConcat}
	\mathbf{M} = \operatorname{concat}( \mathbf{Z}\  || \   \operatorname{concat}(\mathbf{X}\ || \  (\mathbf{V} + \mathbf{A}) ) )
\end{equation}

The matrix $\mathbf{M} \in \mathbb{R}^{1:M \times P}$, as shown in \autoref{eq:FrameConcat}, represents the frame-wise feature matrix from frame $1$ to frame $M$, where each frame has dimensionality $P$, which is the sum of all concatenated feature vectors. Given $\mathbf{X} \in \mathbb{R}^{1:M \times 256}$, $\mathbf{Z} \in \mathbb{R}^{1:M \times 256}$, and $\mathbf{A}, \mathbf{V} \in \mathbb{R}^{1:M \times 64}$, the resulting dimensionality is $P = 256 + 256 + 64$.

Subsequently, the matrix $\mathbf{m}$ is passed through a linear transformation to reduce its dimensionality from $P$ to $256$, resulting in the matrix $\mathbf{m}_{t} \in \mathbb{R}^{1:M \times 256}$.

\begin{equation}
	\label{eq:FeatureDimensionReducion}
	\mathbf{m}_{t} = \operatorname{Linear}_{\theta}( \mathbf{M} )
\end{equation}

\subsubsection{Attention Mechanism in the Feature Integration Process}

In the proposed model, the attention mechanism \cite{vaswani2017attention} is employed to integrate features. The objective of applying attention is to capture the correlation between individual frames in the sequence. The attention mechanism is utilized in both the Cross-Local Attention and the Self-Attention layers within the Transformer Encoder.

The attention mechanism is formulated as follows:

\begin{equation} \label{eq:attention}
	\operatorname{Attention}(\mathbf{Q}, \mathbf{K}, \mathbf{V}, \mathbf{Mask})=\operatorname{softmax}\left(\frac{\mathbf{Q} \mathbf{K}^{T}+\mathbf{Mask}}{\sqrt{C}}\right) \mathbf{V}
\end{equation}

In the Attention formula above, $\mathbf{Q}$ (Query), $\mathbf{K}$ (Key), and $\mathbf{V}$ (Value) are matrices obtained by passing the input through linear transformation matrices: $\mathbf{Q} = {X} \mathbf{W}_Q$, $\mathbf{K} = {X} \mathbf{W}_K$, $\mathbf{V} = {X} \mathbf{W}_V$. The input $X$ is a matrix representing a sequence of $M$ frames, where each frame is a concatenated vector composed of various feature vectors, including seed gestures, text, speech, emotion, and the gesture $\bx_t$ that we aims to denoise. The term $\sqrt{C}$ is a normalization \textbf{c}onstant that accounts for the dimensionality of the matrices. 

The Local-Cross Attention mechanism is controlled to focus only on local motion features of gestures and features in neighboring frames.

\begin{figure}[h]
	\centering
	\includegraphics[width=1\linewidth]{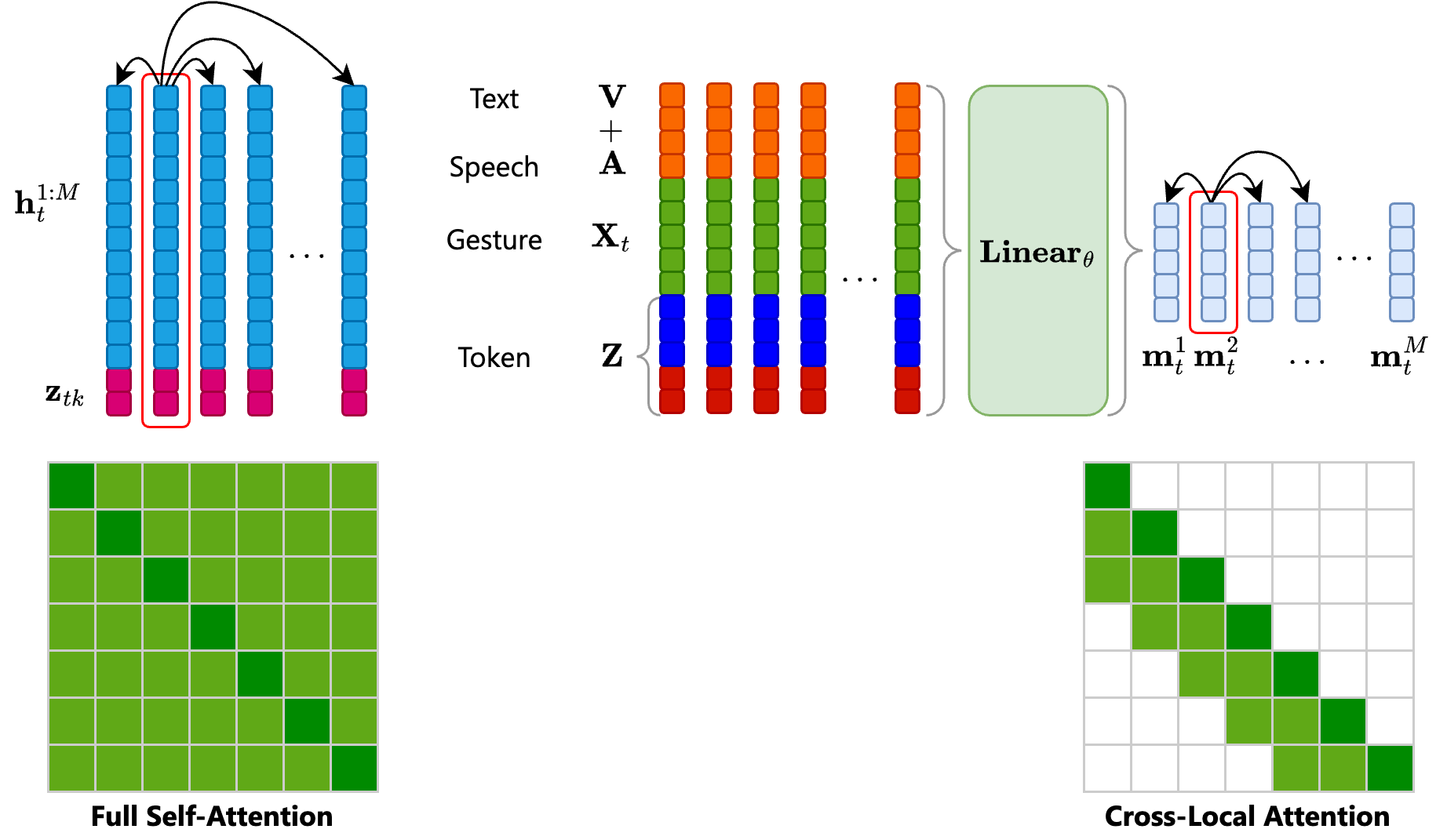}
	\caption{Attention Mechanism in Transformer Encoder and Cross-Local Attention}
	\label{fig:CrossLocalAttention}
\end{figure}

The Attention mechanism functions like a dictionary, where the final retrieved information is the matrix $\mathbf{V}$ (value), while $\mathbf{Q}$ (query) represents the keyword being searched for, and $\mathbf{K}$ (key) is the set of keywords in the lookup dictionary. The Attention process computes the similarity between \( \mathbf{Q} \) and \( \mathbf{K} \) to determine the weights for the values in \( \mathbf{V} \).

The final result is a weighted combination of the values in \( \mathbf{V} \), where the values corresponding to keys most similar to the query receive higher weights. $\mathbf{M}$ is the mask used to perform local attention. The Cross-Local Attention mechanism is illustrated on the right in \autoref{fig:CrossLocalAttention}, while the Transformer Encoder layer uses Self-Attention shown on the left.

\subsubsection{Combining Local Features with Cross-Local Attention}


The matrix $\mathbf{m}_{t} \in \mathbb{R}^{1:M \times 256}$ is then passed through the Cross-Local Attention layer to compute the correlations between local features.

\begin{equation}
	\mathbf{h}_{t}  = \operatorname{Linear}_{\theta}  ( \operatorname{Cross-Local\ Attention}( \mathbf{m}_{t}) )
	\label{eq:CrossLocalAttention}
\end{equation}

Cross-local Attention is performed with $\mathbf{Q} = \mathbf{K} = \mathbf{V} = \mathbf{m}_{t}$.
Following the idea of the Routing Transformer method \cite{roy2021efficient}, Cross-local Attention highlights the importance of constructing intermediate feature-vector representations before passing them through the Transformer Encoder layer, as shown in \autoref{fig:OHGesture}.
The feature vectors are augmented with a relative position-encoding vector \textbf{RPE} (\textbf{R}elative \textbf{P}osition \textbf{E}ncoding) to preserve temporal ordering before entering the Cross-Local Attention layer.

After Cross-Local Attention, the model is forwarded through a linear layer, as in \autoref{eq:CrossLocalAttention}, to align with the $M$ frames and obtain the matrix $\mathbf{h}_t \in \mathbb{R}^{1:M \times D}$.

\subsubsection{Global Feature Fusion with the Transformer Encoder}

Following MDM \cite{tevet2022human}, the vector $\mathbf{z}_{tk}$ is the first token that encodes information for the entire frame sequence, analogous to the $\texttt{CLS}$ token in BERT \cite{devlin2019bertpretrainingdeepbidirectional}, which summarizes an entire text segment.
Here, we use $\mathbf{z}_{tk}$, with $\mathbf{z}_{tk} \in \mathbb{R}^{256}$ (\autoref{eq:ConditionConcat}), as the first token representing global features for the whole sequence of $M$ frames.

\begin{equation}
	\mathbf{X}_{0} = \operatorname{Transformer\ Encoder}\bigl(\operatorname{concat}(\mathbf{z}_{tk} \;\|\; \mathbf{h}^{1:M}_{t})\bigr)
	\label{eq:TransformerEncoder}
\end{equation}

The vectors $\mathbf{h}_t$ represent the sequence of $M$ frames. Similar to Reformer \cite{kitaev2020reformer}, before entering the Self-Attention layer of the Transformer Encoder, the model employs Relative Position Encoding (RPE) instead of absolute position encoding, improving efficiency on long sequences.
Within the Transformer Encoder layer \cite{vaswani2017attention}, relationships among the data sequences are computed.
The Transformer Encoder applies the same self-attention mechanism as in \autoref{eq:attention} but without the $\mathbf{Mask}$, enabling correlations across the entire sequence to be captured.

\subsection{Feature Decoding Stage}

In stage \textit{6. Feature Decoding} (\autoref{fig:CommonStage}), once feature correlations are computed, the goal is to upsample the data back to its original dimensionality.

As illustrated in \autoref{fig:OHGesture}, the latent matrix $\mathbf{X}_{0}$, after passing through the Transformer Encoder to capture correlations among heterogeneous data types, is fed into a linear projection layer
$\hat{\mathbf{x}}_{0} = \operatorname{Linear}_{\theta}(\mathbf{X}_{0})$
to restore the latent matrix to its original size, yielding $\hat{\mathbf{x}}_{0} \in \mathbb{R}^{1:M \times D}$.

The final rendering step is presented in \autoref{sec:Render}.

\subsection{Emotion Control in Gesture Generation}

The preceding steps enable the model to learn gesture generation. To incorporate emotions across different contexts, each emotion is parameterized and varied so that the predictions faithfully express the designated affect.

Analogous to conditional denoising models \cite{ho2022classifier, tevet2022human}, we use the condition vector
$c = [\mathbf{s}, \mathbf{e}, \mathbf{a}, \mathbf{v}]$,  
where $\mathbf{s}$ is the seed gesture, $\mathbf{e}$ the emotion, $\mathbf{a}$ the associated speech, and $\mathbf{v}$ the text.
The conditional diffusion model injects $c$ at every timestep $t$ in the denoising network $\text{G}_\theta(\bx_{t}, t, c)$, with
$c_{\varnothing} = [\varnothing, \varnothing, \mathbf{a}, \mathbf{v}]$ (unconditional)
and $c = [\mathbf{s}, \mathbf{e}, \mathbf{a}, \mathbf{v}]$ (conditional).
A random mask applied to the seed-gesture and emotion vectors conveniently switches labels, allowing optimization under diverse conditions.

\begin{equation} \label{eq:denoise}
	\hat{\bx}_{0\,c,c_{\varnothing},\gamma}
	= \gamma\, G(\bx_{t}, t, c) + (1-\gamma)\, G(\bx_{t}, t, c_{\varnothing})
\end{equation}

Classifier-free guidance \cite{ho2022classifier} further enables interpolation between two emotions
$\mathbf{e}_1$ and $\mathbf{e}_2$ by setting
$c = [\mathbf{s}, \mathbf{e}_{1}, \mathbf{a}, \mathbf{v}]$ and
$c_{\varnothing} = [\mathbf{s}, \mathbf{e}_{2}, \mathbf{a}, \mathbf{v}]$:
\[
\hat{x}_{0\,\gamma, c_{1}, c_{2}}
= \gamma\, G(x_{t}, t, c_{1}) + (1-\gamma)\, G(x_{t}, t, c_{2}).
\]

\subsection{Training Procedure}

\begin{algorithm}[h]
	\caption{Training in OHGesture}
	\label{alg:trainingDeepGesture}
	\setlength{\baselineskip}{10pt}
	\begin{enumerate}
		\item Pre-compute $\gamma$, $\sqrt{\alpha_t}$, $\sqrt{1-\alpha_t}$, $\sqrt{\bar{\alpha}_t}$, and random noise $\boldsymbol{\epsilon}_t$ for each timestep $t: 1 \rightarrow T$. Define the noise schedule $\{\alpha_t \in (0,1)\}_{t=1}^T$.
		\item Sample the initial label $\mathbf{x}_0$ from the normalized data distribution.
		\item Randomly generate Bernoulli masks
		$c_{1} = [ \mathbf{s}, \mathbf{e_1}, \mathbf{a}, \mathbf{v} ]$,
		$c_{2} = [ \mathbf{s}, \mathbf{e_2}, \mathbf{a}, \mathbf{v} ]$, or
		$c_{2} = [ \varnothing, \varnothing, \mathbf{a}, \mathbf{v} ]$.
		\item Add noise to obtain the noisy gesture $\mathbf{x}_t$:
		\[
		\mathbf{x}_t = \sqrt{\bar{\alpha}_t}\,\mathbf{x}_0 + \sqrt{1-\bar{\alpha}_t}\,\boldsymbol{\epsilon}_t.
		\]
		\item Sample $t$ \textbf{uniformly} from $[1, T]$.
		\item Given $\mathbf{x}_t$, $t$, and masks $c_1$, $c_2$, predict the gesture sequence:
		\[
		\hat{\mathbf{x}}_{0\,\gamma,c_{1},c_{2}}
		= \gamma\, G_{\theta}(\mathbf{x}_{t}, t, c_{1})
		+ (1-\gamma)\, G_{\theta}(\mathbf{x}_{t}, t, c_{2}).
		\]
		\item Compute the loss and gradient to update $\theta$:
		\[
		\mathcal{L}^t
		= \mathbb{E}_{t, \mathbf{x}_0, \boldsymbol{\epsilon}_t}
		\bigl[\operatorname{HuberLoss}(\mathbf{x}_0, \hat{\mathbf{x}}_0)\bigr].
		\]
		\item Repeat from step 6 until convergence, obtaining the optimal parameters $\theta'$.
	\end{enumerate}
\end{algorithm}

\autoref{alg:trainingDeepGesture} trains the OHGesture model by first computing the required values and hyper-parameters -- 	$\gamma$, $\sqrt{\alpha_t}$, $\sqrt{1-\alpha_t}$, $\sqrt{\bar{\alpha}_t}$, and $\boldsymbol{\epsilon}_t$ -- for every timestep $t$ (1 … $T$).  
The initial label $\mathbf{x}_0$, representing the ground-truth gesture, is drawn from the normalized data distribution.  
Random Bernoulli masks $c_1$ and $c_2$ emulate different conditions (gesture, emotion, speech, or text), with one mask possibly lacking emotion information.  
Noise is then added to create the noisy gesture $\mathbf{x}_t$.  
A timestep $t$ is sampled uniformly, and $\mathbf{x}_t$ with the masks is fed into the model to predict the original gesture sequence as a weighted combination of conditional outputs.  
The Huber loss between ground-truth and prediction is used to update $\theta$.  
This cycle repeats until the model converges, yielding the optimal parameters $\theta'$.

\subsection{Sampling Process}

\begin{figure}[h]
	\centering
	\includegraphics[width=\linewidth]{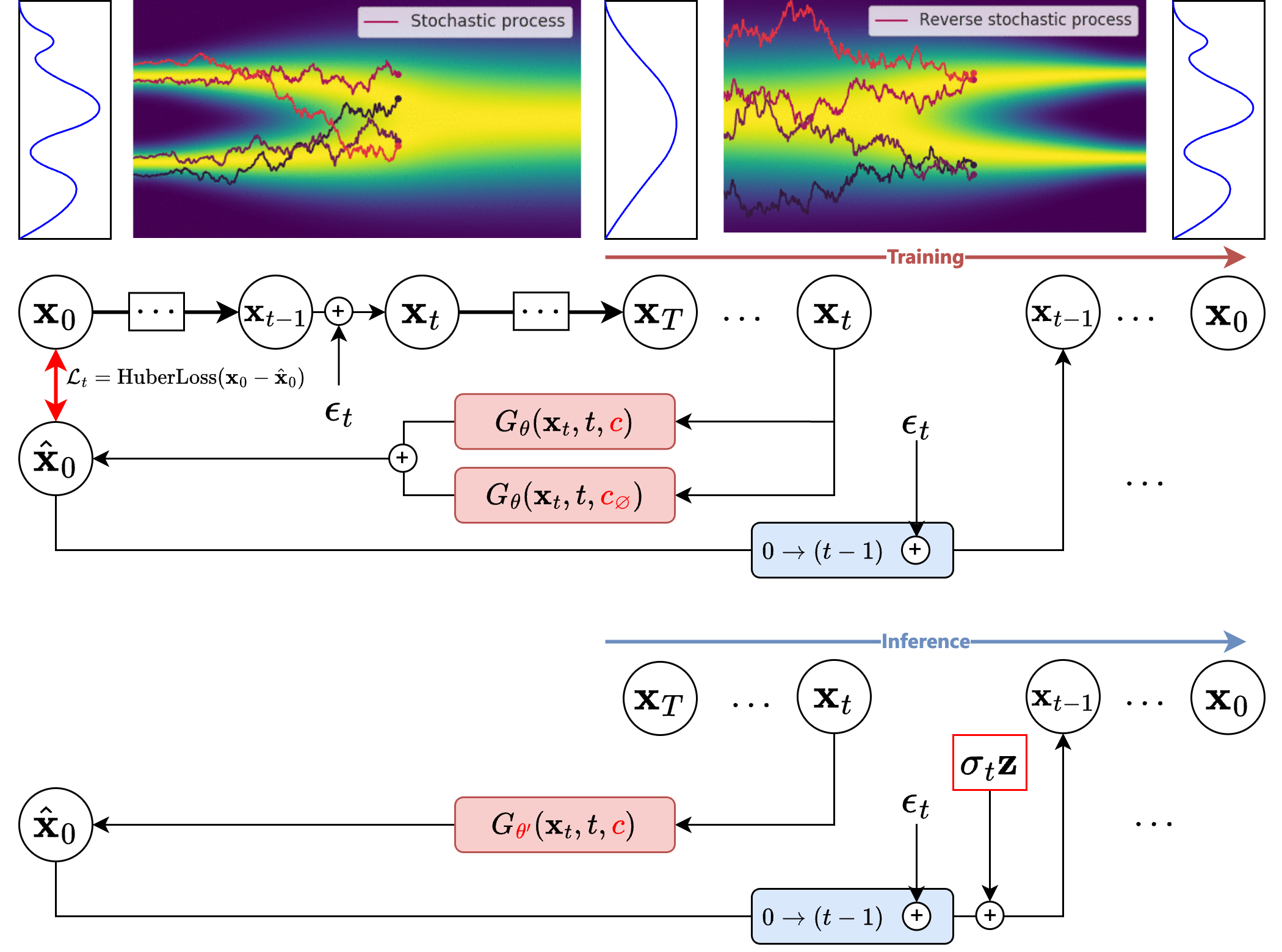}
	\caption{Offline (Training) and Online (Inference) Phases}
	\label{fig:OnlineAndOffline}
\end{figure}

\vspace{-5mm}

To generate gestures of arbitrary length, the original sequence is segmented into clips of length $M$.
During training, the seed gesture can be chosen by randomly selecting a gesture from the dataset or by averaging the clipped segments-here, the mean rotation angles are used.  
Generated frames are processed sequentially, with the last $N=8$ frames taken as the seed for the next iteration.  
For each clip, the gesture $\bx_{t}$ is denoised via $\hat{\bx}_{0} = G_{\theta'}(\bx_{t}, t, c)$; noise is re-added to obtain $\bx_{t-1}$, and the procedure repeats until $t=1$, yielding $\bx_{0}$.

\begin{algorithm}[h]
	\caption{Sampling in OHGesture}
	\label{alg:sampling}
	\setlength{\baselineskip}{10pt}
	\begin{enumerate}
		\item Initialize with noise: $\mathbf{x}_T \sim \mathcal{N}(0, \mathbf{I})$.
		\item Retrieve $\sqrt{\alpha_t}$, $\sqrt{1 - \alpha_t}$, and $\sqrt{\bar{\alpha}_t}$ from training; precompute $\sigma_t$ from $\alpha_t$ for each timestep $t: 1 \rightarrow T$.
		\item Split each 4-second speech segment into $\mathbf{a} \in \mathbb{R}^{64000}$.  
		The initial seed gesture $\mathbf{s}$ is the data mean and later updated from the inferred gesture segment.  
		Select the desired emotion, obtain the transcript $\mathbf{v}$ from speech $\mathbf{a}$, and form the condition $c = [\mathbf{s}, \mathbf{e}, \mathbf{a}, \mathbf{v}]$.
		\item For each timestep, take $t$ \textbf{sequentially} from $[T, \dots, 1]$.
		\item Sample random noise $\mathbf{z} \sim \mathcal{N}(0, \mathbf{I})$.
		\item Infer $\hat{\mathbf{x}}_0^{(t)} = G_{\theta'}(\mathbf{x}_t, t, c)$.
		\item Diffuse $\hat{\mathbf{x}}_0^{(t)}$ from step $0 \rightarrow t$ to obtain $\hat{\mathbf{x}}_{t-1}^{(t)}$.
		\item Add noise: $\hat{\mathbf{x}}_{t-1} = \hat{\mathbf{x}}_{t-1}^{(t)} + \sigma_t \mathbf{z}$.
		\item Return to step 4.  
		When $t = 1$, output the denoised gesture $\hat{\mathbf{x}}_0$.
	\end{enumerate}
\end{algorithm}

\autoref{alg:sampling} starts by initializing the noisy gesture $\mathbf{x}_T$ from $\mathcal{N}(0, \mathbf{I})$.  
The values $\sqrt{\alpha_t}$, $\sqrt{1-\alpha_t}$, and $\sqrt{\bar{\alpha}_t}$ obtained during training, together with $\sigma_t$, are employed at each timestep (1 … $T$).  
Each 4-second speech segment is represented by $\mathbf{a}$, and the seed gesture $\mathbf{s}$ is taken as the data mean or from the previously inferred segment.  
The desired emotion and the transcript form the condition $c = [\mathbf{s}, \mathbf{e}, \mathbf{a}, \mathbf{v}]$.  
The algorithm proceeds sequentially from $T$ to 1: random noise $\mathbf{z}$ is generated, the model predicts $\hat{\mathbf{x}}_0^{(t)}$ from $\mathbf{x}_t$, $t$, and $c$, then $\hat{\mathbf{x}}_{t-1}^{(t)}$ is computed and perturbed with noise.  
This loop continues until $t=1$, after which the algorithm outputs the final denoised gesture $\hat{\mathbf{x}}_0$.

%% file: sections/4_data_preparation.tex
\section{EXPERIMENTS AND RENDERING}
\label{sec:results}

\subsection{Dataset}


\begin{figure}[htbp]
	\centering
	\includegraphics[width=\linewidth]{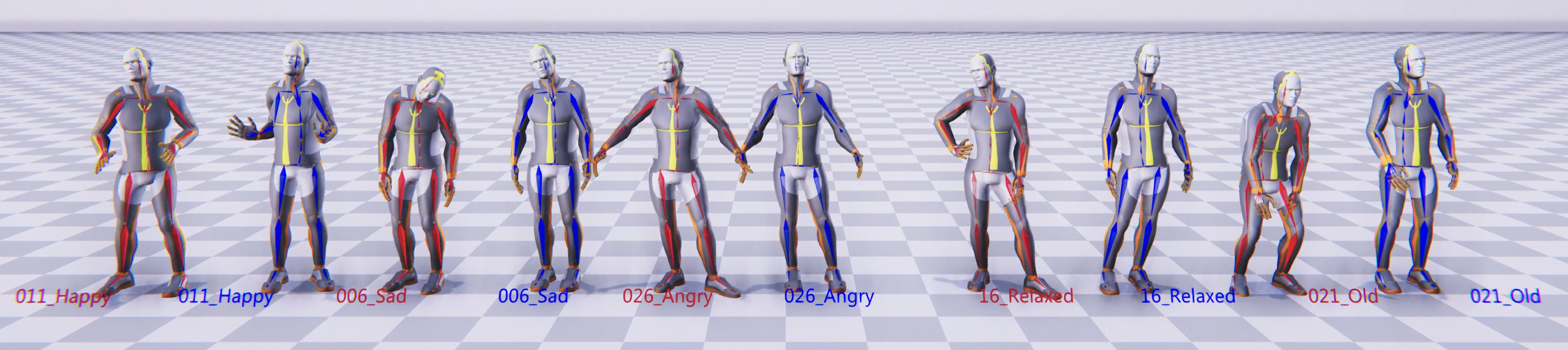}
	\caption{Gesture samples of difference emotion}
\end{figure}

We uses the ZeroEGGS retarget dataset \cite{ghorbani2022zeroeggs}, \cite{ghorbani2022zeroeggszeroshotexamplebasedgesture}, a motion capture dataset designed for research and development of gesture generation models. It includes 67 monologue clips performed by a female motion capture actor, with a total duration of 135 minutes. The monologues are annotated with 6 different emotions: $\texttt{Happy}$, $\texttt{Sad}$, $\texttt{Neutral}$, $\texttt{Old}$, $\texttt{Relaxed}$, and $\texttt{Angry}$, enabling simulation of various emotional states in gestures and body movements. ZeroEGGS provides a rich platform for studying the integration of speech and dynamic gestures, supporting the development of models capable of adapting gestures according to emotions and text semantics.

\subsection{Data Preprocessing}
\label{sec:Preprocessing}

In stage {1. Data Preprocessing} (\autoref{fig:CommonStage}), gesture, speech, and text data are read and processed to be represented as vectors or matrices containing information derived from raw data.

For \textbf{text data}: we use the $\texttt{nltk}$ library for tokenization and $\texttt{contractions}$ to normalize contracted words.

One of our contributions is converting the speech data available in ZeroEGGS using Adobe Speech To Text, then align the phonetic timestamps using the Montreal Forced Aligner \cite{saxon2020robust} with an English phoneme dictionary to match the gesture frame rate, generating TextGrid files. From these TextGrids containing word-level timing information, we use $\texttt{gensim}$ to generate word2vec embeddings.

\textbf{Gesture data} consists of BVH (BioVision Motion Capture) files captured via motion capture systems. BVH files include two components: Hierarchy and Motion. Specifically:

\begin{itemize}
	\item \texttt{HIERARCHY}: defines a skeletal tree containing 75 bones $\{ \mathbf{b}_1, \mathbf{b}_2 \cdots \mathbf{b}_{75} \}$, each with an initial position (\texttt{OFFSET}) and \texttt{CHANNELS} parameters specifying the type and order of rotation angles (\texttt{Zrotation}, \texttt{Yrotation}, \texttt{Xrotation}) and position (\texttt{Xposition}, \texttt{Yposition}, \texttt{Zposition}), which are defined in the \texttt{MOTION} section. The first bone (usually \texttt{Hips}) is the root bone $\mathbf{b}_{\text{root}}$, used to define the T-pose via forward kinematics as the initial pose of the skeleton before applying motion.
	
	\item \texttt{MOTION}: a sequence of frames, each containing motion data representing changes of all $75$ bones as defined by the \texttt{CHANNELS} in the \texttt{HIERARCHY}.
\end{itemize}

Our model converts Euler rotation angles to quaternion rotation angles, where a quaternion is a 4-dimensional vector.

\begin{equation} \label{eq:gesturevector}
	\mathbf{g} = \Big[ \mathbf{p}_{\text{root}},  \mathbf{r}_{\text{root}},
	\mathbf{ p }'_{\text{root}},  \mathbf{r}'_{\text{root}},
	\mathbf{p}_{\text{joins}},  \mathbf{r}_{\text{joins}},
	\mathbf{p}'_{\text{joins}},  \mathbf{r}'_{\text{joins}},
	\mathbf{d}_{\text{gaze}}
	\Big]
\end{equation}

Here, each $\mathbf{g} \in \mathbb{R}^{1141}$ includes:
{
	\begin{itemize}
		\item $\mathbf{p}_{\text{root}} \in \mathbb{R}^3$: coordinates of the root joint
		\item $\mathbf{r}_{\text{root}} \in \mathbb{R}^4$: rotation (quaternion) of the root joint
		\item $\mathbf{p}'_{\text{root}} \in \mathbb{R}^3$: velocity of the root position
		\item $\mathbf{r}'_{\text{root}} \in \mathbb{R}^3$: angular velocity of the root rotation
		
		\item $\mathbf{p}_{\text{joins}} \in \mathbb{R}^{3 n_{\text{join} }}$: positions of other joints
		\item $\mathbf{r}_{\text{joins}} \in \mathbb{R}^{6 n_{\text{join} }}$: joint rotations on the X and Y planes
		\item $\mathbf{p}'_{\text{joins}} \in \mathbb{R}^{3n_{\text{join} }}$: velocity of joint positions
		\item $\mathbf{r}'_{\text{joins}} \in \mathbb{R}^{3n_{\text{join} }}$: angular velocity of joint rotations
		\item $\mathbf{d}_{\text{gaze}} \in \mathbb{R}^3$: gaze direction
\end{itemize}}

The original gesture sequences in Euler angles are converted to radians, then converted from Euler to Quaternion as detailed in \autoref{appendix:BVHQuaternion}.

\textbf{Speech data}: $\mathbf{a}_{\text{raw}} \in \mathbb{R}^{ \text{length } }$ is raw speech sampled at 16000 Hz, trimmed into segments $\mathbf{a} \in \mathbb{R}^{64000}$ corresponding to 4 seconds. The paper uses \texttt{ffmpeg-normalize} to normalize volume to a level lower than the original.

\textbf{Emotion}: Emotion data is represented using a predefined one-hot encoded vector. During sampling, the filename encodes the target emotion.

All data is stored using the $\texttt{h5}$ format.

\subsection{Training Process}

The entire model training process was conducted over approximately two weeks with the following parameters: number of training steps $T = 1000$, using an Nvidia 3090 GPU. The learning rate was set to $3 \times 10^{-5}$, batch size was $640$, and a total of $43,853$ samples were trained. At each step, $t$ is randomly sampled and input to $f_{\theta}$ to predict $\mathbf{x}_{0}$. The emotional control parameter was set to $\gamma = 0.1$. The probability of applying random masking to the emotion and initial gesture matrices was $10\%$, using a Bernoulli distribution to randomly hide/reveal these matrices.

The $\beta$ parameter was scheduled linearly from $0.5 \rightarrow 0.999$.

The $\operatorname{HuberLoss} (\mathbf{x}_{0},  \hat{\mathbf{x}}_{0} )$ is computed as follows:

\begin{itemize}
	\item If $|\mathbf{x}_0 - \hat{\mathbf{x}}_0| \leq \delta$ then $\mathcal{L}_{ \delta, \mathbf{x}_0, \hat{\mathbf{x}}_0} = \frac{1}{2} (\mathbf{x}_0 - \mathbf{x}_0)^2$: Below the threshold, the loss is computed as squared distance (similar to MSE), which is sensitive to small errors and provides smooth gradients.
	
	\item If $|\mathbf{x}_0 - \hat{\mathbf{x}}_0| > \delta$ then $\mathcal{L}_{ \delta, \mathbf{x}_0, \hat{\mathbf{x}}_0}  =  \delta \cdot |\mathbf{x}_0 - \mathbf{x}_0| - \frac{1}{2} \delta^2$: Above the threshold, the loss behaves like MAE, reducing sensitivity to large errors and improving robustness against outliers.
	
\end{itemize}

The training process is implemented in the open-source repository: \hyperlink{https://github.com/OHGesture/OHGesture}{Github/OHGesture} \footnote{\url{https://github.com/OHGesture/OHGesture}}.

\subsection{Rendering Process in Unity}
\label{sec:Render}

To visualize the gesture generation process from model output, we use Unity in stage \textit{7. Rendering} (\autoref{fig:CommonStage}), extending code from the DeepPhase model \cite{starke2022deepphase}. The generated output is in BVH (BioVision Motion Capture) format. In Unity, the author adds C-Sharp scripts to render gestures based on coordinates and labels, with bone positions and rotations represented using quaternions.

Rendering details are presented in \autoref{appendix3}.

The Unity project source code is available at: \hyperlink{https://github.com/DeepGesture/deepgesture-unity}{Github/DeepGesture-Unity}
\footnote{\url{https://github.com/DeepGesture/deepgesture-unity}}.

%% file: sections/7_results.tex
\section{EVALUATION AND RESULTS}
\label{chap:evalution}

\subsection{Evaluation Methods}

The evaluation is conducted using two primary metrics: Mean Opinion Scores (MOS) and Fréchet Gesture Distance (FGD).

\subsubsection{Human Perception Evaluation}

\textbf{Mean Opinion Scores (MOS)}

As of now, there is no standardized metric for gesture generation, particularly in the speech-driven context. Consequently, this work relies on subjective human evaluations, a practice consistent with prior research~\cite{yoon2022genea, kucherenko2021large, alexanderson2022listen}, since objective metrics still fail to fully capture human perception.

MOS is assessed based on the following three criteria:

\begin{itemize}
	\item Human-likeness
	\item Gesture-Speech Appropriateness
	\item Gesture-Style Appropriateness
\end{itemize}

A notable contribution of this work is the development of the \hyperlink{https://genea-workshop.github.io/leaderboard/}{GENEA Leaderboard}~\cite{nagy2024towards}, which includes HEMVIP (\textbf{H}uman \textbf{E}valuation of \textbf{M}ultiple \textbf{V}ideos in \textbf{P}arallel). HEMVIP is designed to compare gesture generation quality between different models using rendered videos.

Within the GENEA consortium (\textbf{G}eneration and \textbf{E}valuation of \textbf{N}on-verbal Behaviour for \textbf{E}mbodied \textbf{A}gents), evaluators are recruited via Prolific. Participants watch videos rendered by different models and assign ratings of \textit{Left Better}, \textit{Equal}, or \textit{Right Better}. These responses are converted into numerical scores of $-1$, $0$, and $1$, respectively, and updated via the Elo rating system.

The evaluation platform and source code are available at \hyperlink{https://github.com/hemvip/hemvip.github.io/}{github.com/hemvip/hemvip.github.io}\footnote{HEMVIP 2: \url{https://github.com/hemvip/hemvip.github.io}}.

\subsubsection{Objective Metric Evaluation}

\textbf{Mean Square Error (MSE)}

We compute the mean square error between the predicted gesture sequence $\hat{\mathbf{y}}_i^{1:M \times D}$ and the corresponding ground-truth sequence $\mathbf{y}_i^{1:M \times D}$ as:

\begin{equation}
	\text{MSE} = \frac{1}{n} \sum_{i=1}^n \left\| \mathbf{y}_i^{1:M \times D} - \hat{\mathbf{y}}_i^{1:M \times D} \right\|^2
\end{equation}

Where:
\begin{itemize}
	\item $n$ is the number of data samples,
	\item $M$ is the number of frames per sample,
	\item $D$ is the number of dimensions per frame,
	\item $\|\cdot\|^2$ is the squared Frobenius norm.
\end{itemize}

A lower MSE indicates a smaller difference between predicted and ground-truth gestures. Evaluation results are provided in \autoref{subsec:MSEResult}.

\textbf{Fréchet Gesture Distance (FGD)}

Inspired by the Fréchet Inception Distance (FID) used in image generation, the Fréchet Gesture Distance (FGD) measures the similarity in distribution between predicted gestures $\hat{\mathbf{y}}$ and real gestures $\mathbf{y}$:

\begin{equation}
	\text{FGD} = \left\| \hat{\mu} - \mu \right\|^2 + \operatorname{Tr}\left( \Sigma + \hat{\Sigma} - 2 \sqrt{\Sigma \hat{\Sigma}} \right)
	\label{eq:fidscore}
\end{equation}

Where:
\begin{itemize}
	\item $\mu$ and $\hat{\mu}$ are the empirical means of real and generated data,
	\item $\Sigma$ and $\hat{\Sigma}$ are the corresponding covariance matrices,
	\item $\operatorname{Tr}(\cdot)$ denotes the trace operator,
	\item $\sqrt{\Sigma \hat{\Sigma}}$ is the matrix square root of the product.
\end{itemize}

Lower FGD indicates higher distributional similarity, thus better quality. Evaluations are conducted on $\hat{\mathbf{x}}^{0}, \mathbf{x}_0 \in \mathbb{R}^{1:M \times D}$.

\subsection{Evaluation Results}
\label{sec:result}

\subsubsection{User Study Results}

\textbf{MOS Evaluation}

Due to the high cost and resource requirements of human evaluations, this work reuses the subjective evaluation results from \textbf{DiffuseStyleGesture}~\cite{yang2023diffusestylegesture} as a baseline. The proposed \textbf{OHGesture} model is not included in the subjective study.

To assess visual quality, a separate user study is conducted comparing the proposed method with real motion capture data. Each evaluated clip is 11–51 seconds long (mean: 31.6s), exceeding the typical GENEA clip duration of 8–10s~\cite{yoon2022genea}. Longer clips offer clearer insight into gesture consistency~\cite{yang2022reprgesture}. Ratings range from 5 (\texttt{excellent}) to 1 (\texttt{bad}).

\begin{table}[h]
	\centering
	\begin{tabular}{lcc}
		\hline
		\multicolumn{1}{c}{Model} &
		\begin{tabular}[c]{@{}c@{}}Human\\ Likeness\end{tabular}$\uparrow$ &
		\begin{tabular}[c]{@{}c@{}}Gesture-Speech\\ Appropriateness\end{tabular}$\uparrow$ \\ \hline
		Ground Truth          & 4.15 $\pm$ 0.11          & 4.25 $\pm$ 0.09          \\
		Ours                  & \textbf{4.11 $\pm$ 0.08} & \textbf{4.11 $\pm$ 0.10} \\
		\hline
	\end{tabular}
	\caption{Mean Opinion Score (MOS) evaluation results}
	\label{table:MOSScore}
\end{table}

\subsubsection{Quantitative Results}

\textbf{MSE Evaluation}
\label{subsec:MSEResult}

MSE is computed on gesture sequences across $M$ frames and reported across six emotion categories.

\begin{table}[h]
	\centering
	\resizebox{\linewidth}{!}{%
		\begin{tabular}{lcccccc}
			\hline
			\multicolumn{1}{c}{Emotion} & Neutral & Sad & Happy & Relaxed & Elderly & Angry \\ \hline
			DiffuseStyleGesture  & 75.04 & 51.40 & 110.18 & 130.83     & 116.03    & 78.53     \\
			ZeroEGG              & 136.33 & 81.22 & 290.47 & 140.24     & 102.44    & 181.07     \\
			\textbf{OHGesture}   & 161.22 & 89.58 & 279.95 & 156.93     & 99.86     & 215.24     \\
			\hline
		\end{tabular}%
	}
	\caption{MSE results across six emotion categories}
	\label{table:EvaluationMSE}
\end{table}

\textbf{FGD Evaluation}

This work introduces FGD and implements the open-source tool \hyperlink{https://github.com/GestureScore/GestureScore}{GestureScore}\footnote{\url{https://github.com/GestureScore/GestureScore}}. GestureScore encodes gesture frames via an Inception V3-based model into a $32 \times 32$ latent representation. FGD is then computed via \autoref{eq:fidscore}.

\begin{table}[h]
	\centering
	\resizebox{\linewidth}{!}{
		\begin{tabular}{lcc}
			\hline
			\multicolumn{1}{c}{Model Variant} & FGD on Feature Vectors & FGD on Raw Data \\ \hline
			Ground Truth             & -             & -            \\
			OHGesture (Feature D=1141) & 2.058  & 9465.546     \\
			OHGesture (Rotations)      & 3.513  & 9519.129     \\
			\hline
		\end{tabular}
	}
	\caption{FGD results for OHGesture on $\bx^{1:M \times D}$}
	\label{table:EvalFGD}
\end{table}

\begin{itemize}
	\item \textbf{Feature Vectors:} Skeletons from BVH files are encoded into vectors with $D = 1141$ dimensions (\autoref{eq:gesturevector}).
	\item \textbf{Rotations:} Joint rotations are extracted, resulting in $D = 225$ features ($75$ joints $\times$ $3$ rotation angles).
\end{itemize}

%% file: sections/8_conclusion.tex
\section{STRENGTHS AND LIMITATIONS}

The proposed OHGesture model introduces several key advantages that advance the development of more natural and expressive human-machine interaction systems. Nevertheless, certain limitations remain and represent important directions for future work.

\textbf{Strengths:}

\begin{itemize}
	\item \textbf{High realism:} OHGesture generates gestures that exhibit strong human-likeness, closely mirroring the timing and rhythm of natural speech. The model effectively synchronizes gestures with both the emotional tone and semantic content of speech.
	
	\item \textbf{Robust generalization:} Leveraging the denoising capabilities of diffusion models, OHGesture can generate gestures for speech and emotional contexts not encountered during training, indicating strong potential for real-world deployment across varied domains.
	
	\item \textbf{Multidimensional controllability:} The diffusion-based architecture supports flexible control over multiple attributes, including the interpolation and modulation of emotional states, enabling expressive and context-sensitive gesture synthesis.
\end{itemize}

\textbf{Limitations:}

\begin{itemize}
	\item \textbf{Lack of real-time inference:} The current implementation is not optimized for real-time execution and requires multi-step generation followed by offline rendering.
	
	\item \textbf{Suboptimal motion representation:} The gesture feature representation with $D=1141$ dimensions is processed as a 2D image structure, which may not fully capture temporal motion dynamics and biomechanical features.
	
	\item \textbf{Sensitivity to input quality:} The model’s performance heavily depends on clean, high-quality speech input. In cases of noisy or emotionally ambiguous audio, the gesture output may be less accurate or less expressive.
\end{itemize}

\section{CONCLUSION}
\label{sec:conclusion}

\subsection{Achieved Results}

This work presents OHGesture, a gesture generation model that achieves high realism and naturalness through diffusion-based modeling. A primary contribution is its precise synchronization between generated gestures and the emotional content of speech, extending generalization beyond the training distribution. The model demonstrates the ability to produce context-aware gestures even in low-probability speech scenarios.

Another major advancement lies in expanding input modalities. In addition to speech, gesture, and emotion labels, the model incorporates textual features obtained through automatic transcription. This multimodal approach enables the system to capture the semantic context of input speech more effectively and generate more appropriate gestures.

\subsection{Future Work}

Several promising directions can further improve OHGesture and expand its applicability in real-world settings:

\begin{itemize}
	\item \textbf{Real-time inference optimization:} Future efforts will focus on transforming OHGesture into a real-time gesture generation system by reducing dependencies on offline rendering tools (e.g., Unity) and optimizing latency for interactive applications.
	
	\item \textbf{Efficient sampling:} The current diffusion process involves many sampling steps. Reducing these without compromising generation quality remains an important area for improving system responsiveness.
	
	\item \textbf{Advanced embeddings:} Incorporating richer embedding techniques-potentially combining speech, text, prosody, and affective signals-could further enhance gesture appropriateness across different languages and cultural contexts.
	
	\item \textbf{Multilingual generalization:} Extending the model to support multiple languages will broaden its applicability and ensure culturally appropriate gesture behavior.
	
	\item \textbf{Integration with phase-aware models:} A future integration with the DeepPhase model~\cite{starke2022deepphase} is planned to improve the representation of temporal dynamics and support real-time applications via phase-informed motion generation.
	
	\item \textbf{Improved automatic metrics:} To reduce reliance on subjective human evaluation, ongoing work aims to design robust automatic evaluation metrics that better reflect human perception and can serve as internal feedback mechanisms during training and inference.
\end{itemize}

\subsection{Closing Remarks}

Through experimental validation and qualitative analysis, the OHGesture model -- extending DiffuseStyleGesture-demonstrates the ability to generate realistic gestures for both in -- distribution and out-of-distribution speech, including synthetic voices such as that of Steve Jobs (see \autoref{appendix3}). This highlights the promise of diffusion-based approaches in modeling complex, expressive, and low-frequency gesture behaviors.

Additionally, we contribute open-source code, including rendering pipelines and data processing tools built on Unity, available on GitHub. These resources provide a solid foundation for future development and reproducibility. The integration of text alongside speech and emotion into the gesture generation pipeline marks an important step toward building fully multimodal agents, capable of more intuitive and human-like interactions in diverse application domains.

%% file: sections/x_ack.tex
\section*{Acknowledgments}

This research is partially supported by OpenHuman AI. We thank Daniel Holden for providing the retargeting dataset \cite{ghorbani2022zeroeggs}. We also acknowledge the use of the Ubisoft La Forge ZeroEGGS Animation Dataset, which includes a version exported in FBX and BVH formats.

%% file: sections/appendix1.tex
\appendix
\section{BVH Data Processing Pipeline}
\label{appendix:BVHData}

\subsection{Skeleton Structure of a Character}
\label{appendix:BVHSkeleton}

Some skeleton joint names from the $75$ motion skeletons include:

{
	\small
	\texttt{Hips},
	\texttt{Spine},
	\texttt{Neck},
	\texttt{Head},
	\texttt{RightShoulder},
	\texttt{RightArm},
	\texttt{RightForeArm},
	\texttt{RightHand},
	\texttt{LeftShoulder},
	\texttt{LeftArm},
	\texttt{LeftForeArm},
	\texttt{LeftHand},
	\texttt{RightUpLeg},
	\texttt{RightLeg},
	\texttt{RightFoot},
	\texttt{RightToeBase},
	\texttt{LeftUpLeg},
	\texttt{LeftLeg},
	\texttt{LeftFoot},
	\texttt{LeftToeBase},
	...
}

\begin{figure}[h]
	\centering
	\includegraphics[height=10.5cm]{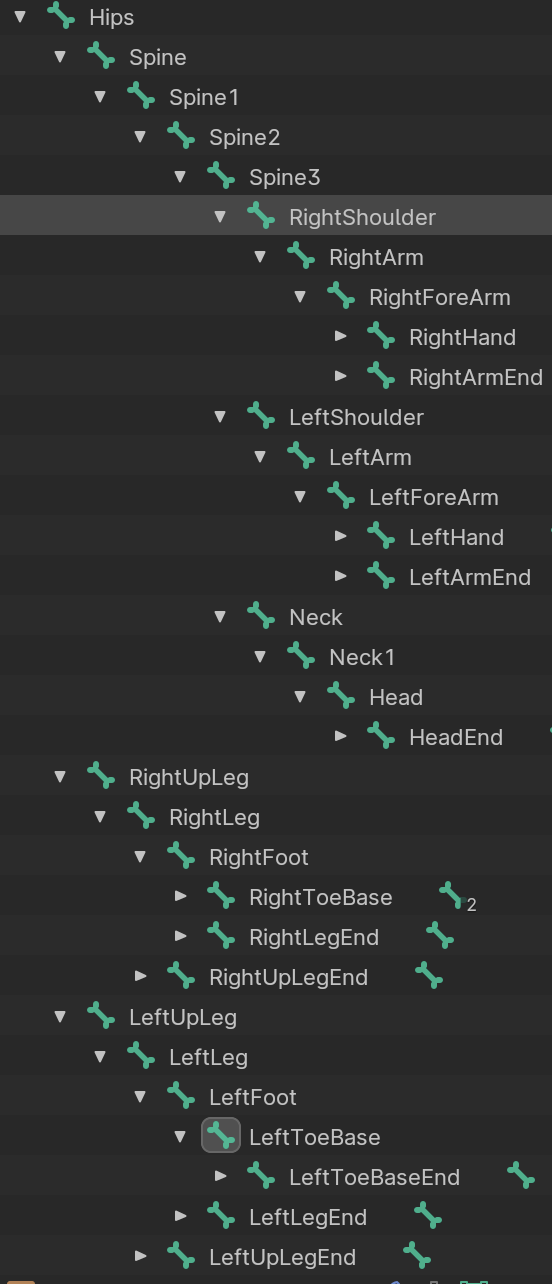}
	\caption{Character skeleton}
	\label{fig:Bone}
\end{figure}

\subsection{Structure of a BVH File}
\label{appendix:BVHStructure}

A BVH (Biovision Hierarchy) file is a data format that contains information about the skeleton structure and motion data of bones in a skeletal system. A BVH file consists of two main parts: the skeleton hierarchy declaration and the bone motion data.

\begin{itemize}
	\item \textbf{HIERARCHY}:
	
	\begin{itemize}
		\item Defines the components and names of the skeleton joints, as well as the initial positions of the joints in the T-pose (motion capture actors extend their arms horizontally to form a "T").
		\item Defines the parent-child relationships from the root node to the leaf nodes of the skeleton, typically with the root node being the spine ($\texttt{Spine}$).
		\item Specifies the data to be recorded such as position or rotation angles along $X, Y, Z$ axes of each joint over time.
	\end{itemize}
	
	\item \textbf{MOTION}: A sequence of movements frame by frame, where each frame contains bone movement data as defined in the HIERARCHY section (e.g., rotation angles or positions).
\end{itemize}

\begin{figure}[htbp]
	\centering
	\begin{subfigure}{0.49\linewidth}
		\centering
		\includegraphics[width=\linewidth]{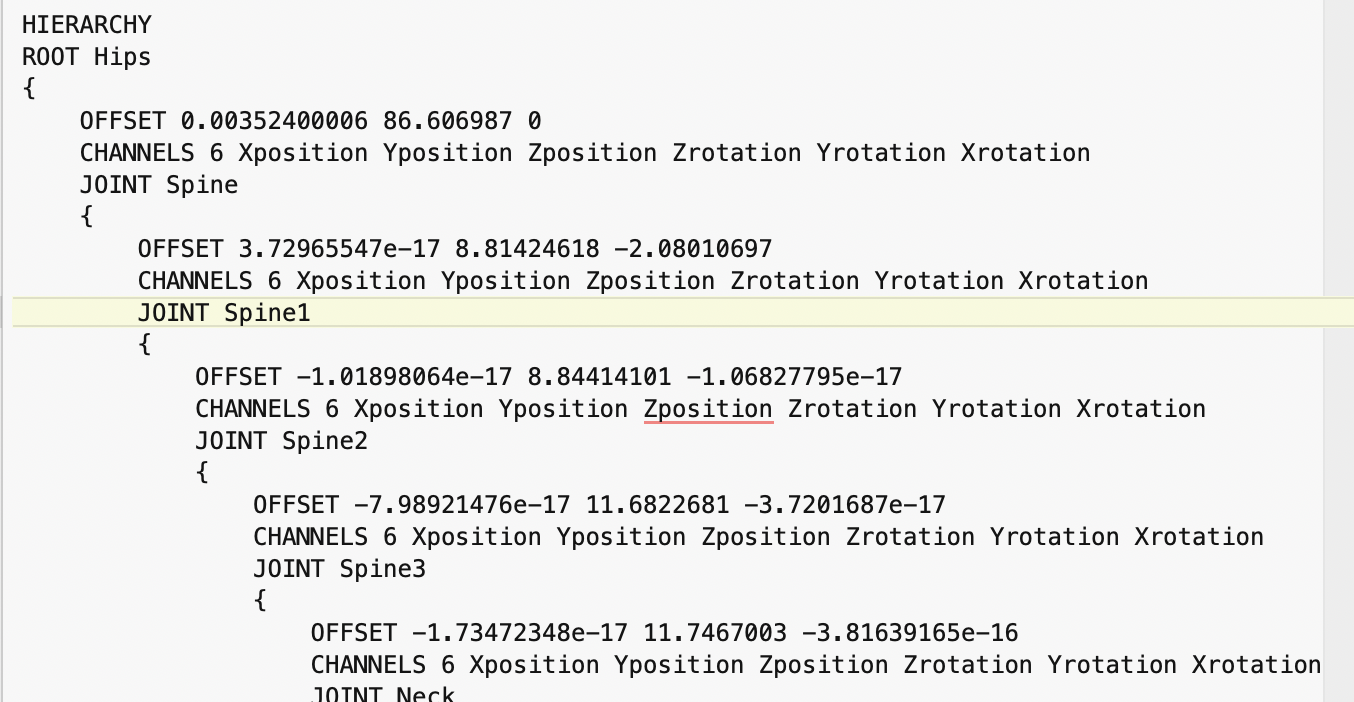}
		\caption{HIERARCHY in BVH file}
		\label{fig:BVH1}
	\end{subfigure}
	\hfill
	\begin{subfigure}{0.49\linewidth}
		\centering
		\includegraphics[width=\linewidth]{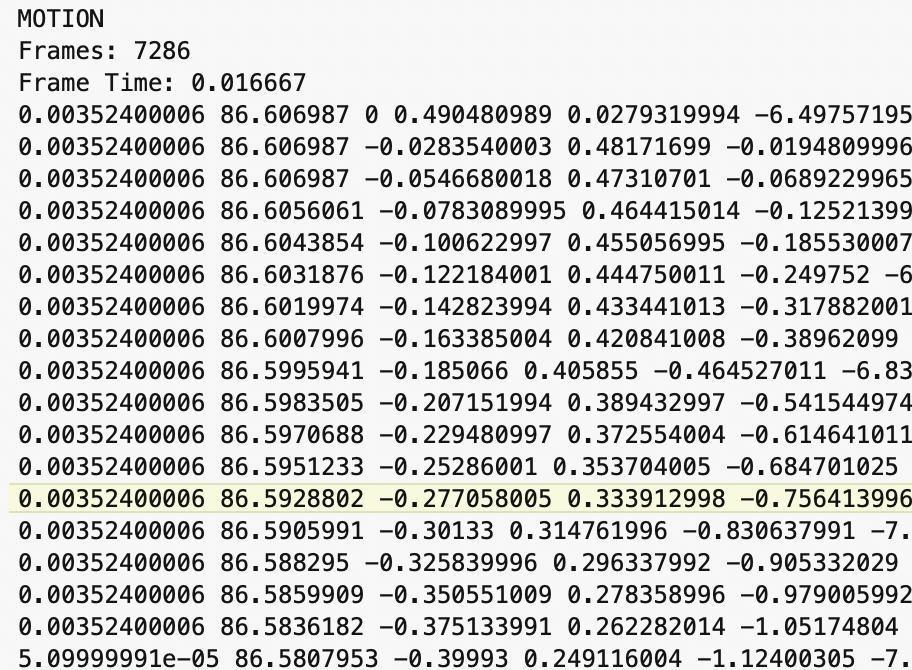}
		\caption{MOTION in BVH file}
		\label{fig:BVH2}
	\end{subfigure}
\end{figure}

$\mathbf{rotation}_i^{\operatorname{local}} = \{ \alpha ,\beta , \gamma \}$ represents the rotation angles around the $Z$, $Y$, and $X$ axes, respectively. The combined rotation in Euler space is:

\begin{equation}
	R = R_Z(\alpha) R_Y(\beta) R_X(\gamma)
\end{equation}

Where:

1. \textbf{Rotation matrix around axis \(Z\)}:

\[
R_Z(\alpha) = 
\begin{bmatrix}
	\cos(\alpha) & -\sin(\alpha) & 0 \\
	\sin(\alpha) & \cos(\alpha) & 0 \\
	0 & 0 & 1
\end{bmatrix}
\]

2. \textbf{Rotation matrix around axis \(Y\)}:

\[
R_Y(\beta) = 
\begin{bmatrix}
	\cos(\beta) & 0 & \sin(\beta) \\
	0 & 1 & 0 \\
	-\sin(\beta) & 0 & \cos(\beta)
\end{bmatrix}
\]

3. \textbf{Rotation matrix around axis \(X\)}:

\[
R_X(\gamma) = 
\begin{bmatrix}
	1 & 0 & 0 \\
	0 & \cos(\gamma) & -\sin(\gamma) \\
	0 & \sin(\gamma) & \cos(\gamma)
\end{bmatrix}
\]

To compute the motion coordinates of a character, the following operation is applied:

\begin{equation}
	\mathbf{position}_{\text{global}} = R \cdot \mathbf{position}_{\text{local}} + \mathbf{t}
\end{equation}

\subsection{Conversion from Euler Angles to Quaternions}
\label{appendix:BVHQuaternion}

To avoid Gimbal lock, Euler angle data must be converted into quaternion representation. Each bone's rotation from Euler angles in the ZYX order is represented as a quaternion $q = (q_w, q_x, q_y, q_z)$, with components calculated as follows:

First, compute the $\cos$ and $\sin$ values of half the rotation angles for each axis:

\begin{itemize}
	\item $c_{\alpha} = \cos\left(\frac{\alpha}{2}\right), \quad s_{\alpha} = \sin\left(\frac{\alpha}{2}\right)$
	\item $c_{\beta} = \cos\left(\frac{\beta}{2}\right), \quad s_{\beta} = \sin\left(\frac{\beta}{2}\right)$
	\item $c_{\gamma} = \cos\left(\frac{\gamma}{2}\right), \quad s_{\gamma} = \sin\left(\frac{\gamma}{2}\right)$
\end{itemize}

Based on the values above, the quaternion components are computed as:

\begin{itemize}
	\item $q_w = c_{\alpha} c_{\beta} c_{\gamma} + s_{\alpha} s_{\beta} s_{\gamma}$
	\item $q_x = c_{\alpha} c_{\beta} s_{\gamma} - s_{\alpha} s_{\beta} c_{\gamma}$
	\item $q_y = c_{\alpha} s_{\beta} c_{\gamma} + s_{\alpha} c_{\beta} s_{\gamma}$
	\item $q_z = s_{\alpha} c_{\beta} c_{\gamma} - c_{\alpha} s_{\beta} s_{\gamma}$
\end{itemize}

With the computed quaternion $q$, the global position of the bone $\mathbf{p}_{\text{global}}$ is determined by rotating the local position $\mathbf{p}_{\text{local}}$ using the formula:

\begin{equation}
	\mathbf{p}_{\text{global}} = q \cdot \mathbf{p}_{\text{local}} \cdot q^{-1} + \mathbf{t}
\end{equation}

where $\mathbf{t}$ is the origin position of the bone in global space.

%% file: sections/appendix2.tex
\section{Illustration of Gesture Inference Results}
\label{appendix3}

\subsection{Comparison between Ground Truth Gestures and Predicted Gestures}

\begin{center}
	\centering
	\href{https://youtu.be/22lNm2tvmrk}{
		\includegraphics[width=\linewidth]{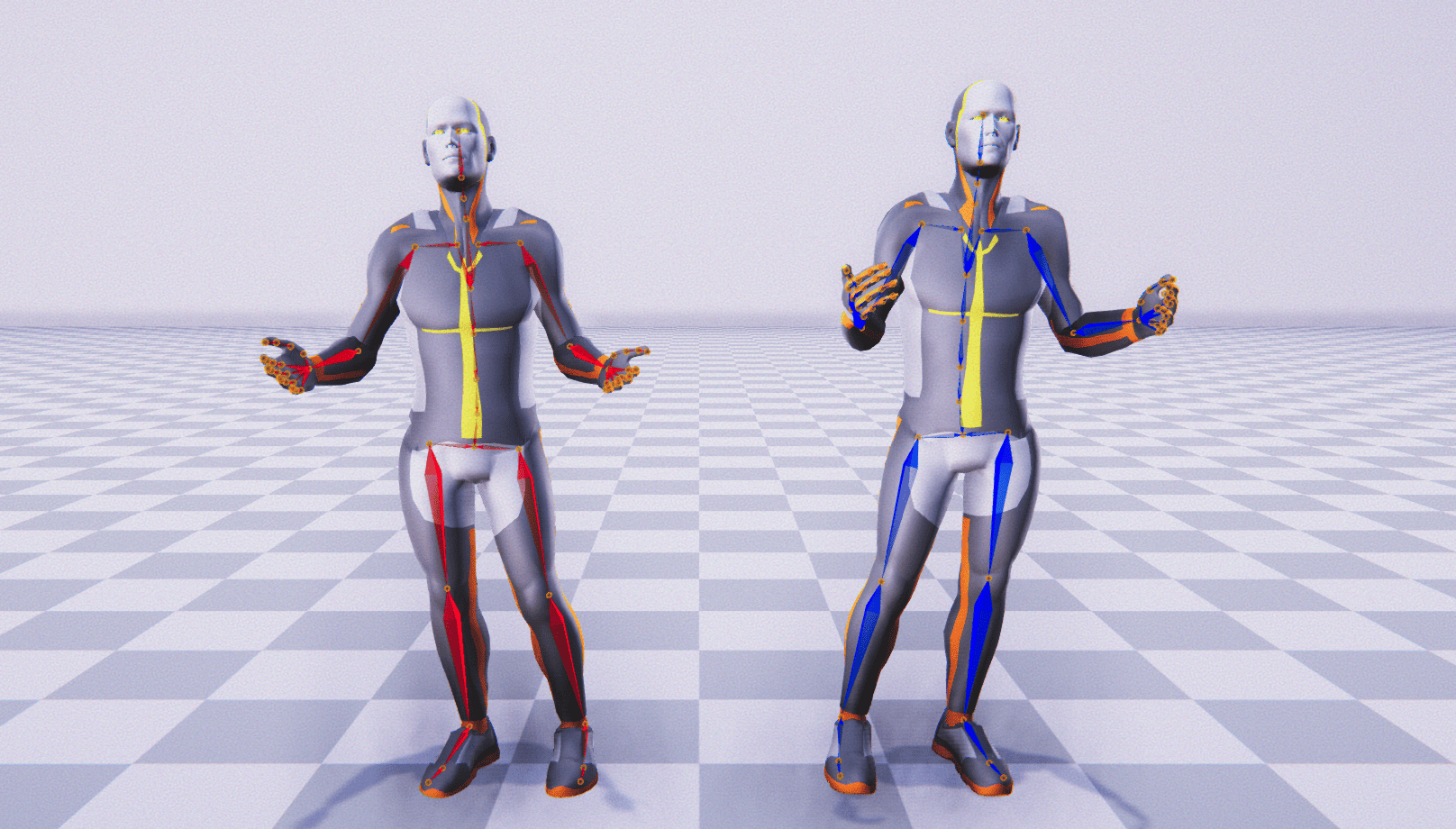}}
	{\tiny Click the image to watch the video}
\end{center}

The results show both the ground truth gestures and the model's predicted gestures at frame $3821$, inferred from the speech sample $\texttt{003\_Neutral\_2\_x\_1\_0}$.

\subsection{Illustration of Different Emotions in Gestures}

{
	\begin{center}
		\centering
		\href{https://youtu.be/KUlBZXLtYJ4}{
			\includegraphics[width=\linewidth]{figures/DifferenceEmotion}}
		{\tiny Click the image to watch the video}
	\end{center}
}

Generated or inferred results from the OHGesture model with different emotions. Red indicates the ground truth from the ZeroEGGS dataset, while blue shows the output generated by the OHGesture model.

{
	\begin{center}
		\centering
		\href{https://youtu.be/eZghfNGmZn8}{
			\includegraphics[width=\linewidth]{figures/ListOfEmotion}}
		
		{\tiny Click the image to watch the video}
	\end{center}
}

\subsection{Illustration of Gesture Generation with Out-of-Training Speech}

{
	\begin{center}
		\centering
		\href{https://www.youtube.com/watch?v=B6nv1kQmi-Q}{
			\includegraphics[width=0.8\linewidth]{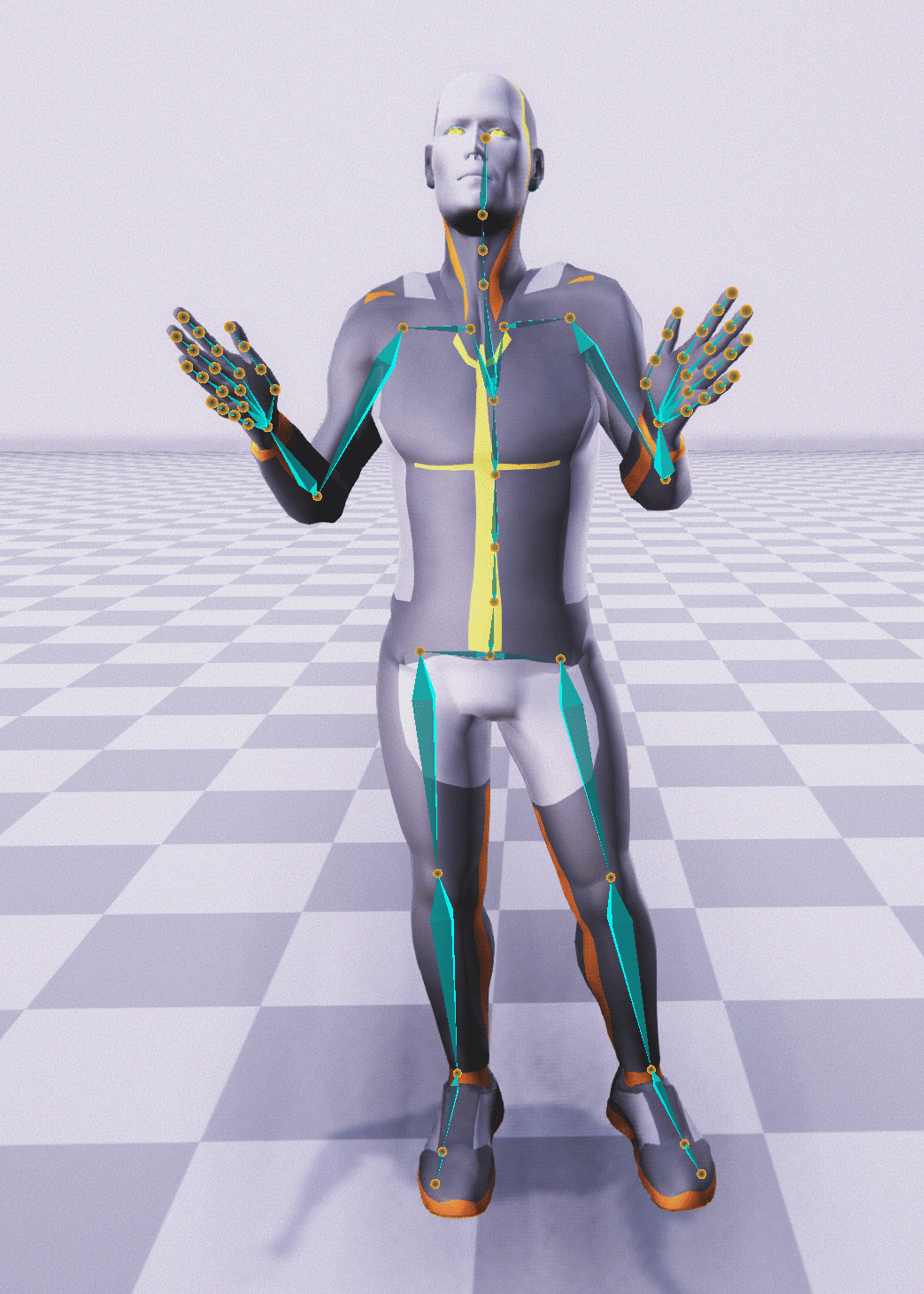}}
		
		{\tiny Click the image to watch the video}
	\end{center}
}

Illustration of gesture generation corresponding to Steve Jobs’ speech.

{
	\begin{center}
		\centering
		\href{https://youtu.be/yLwXdm7UgPE}{
			\includegraphics[width=\linewidth]{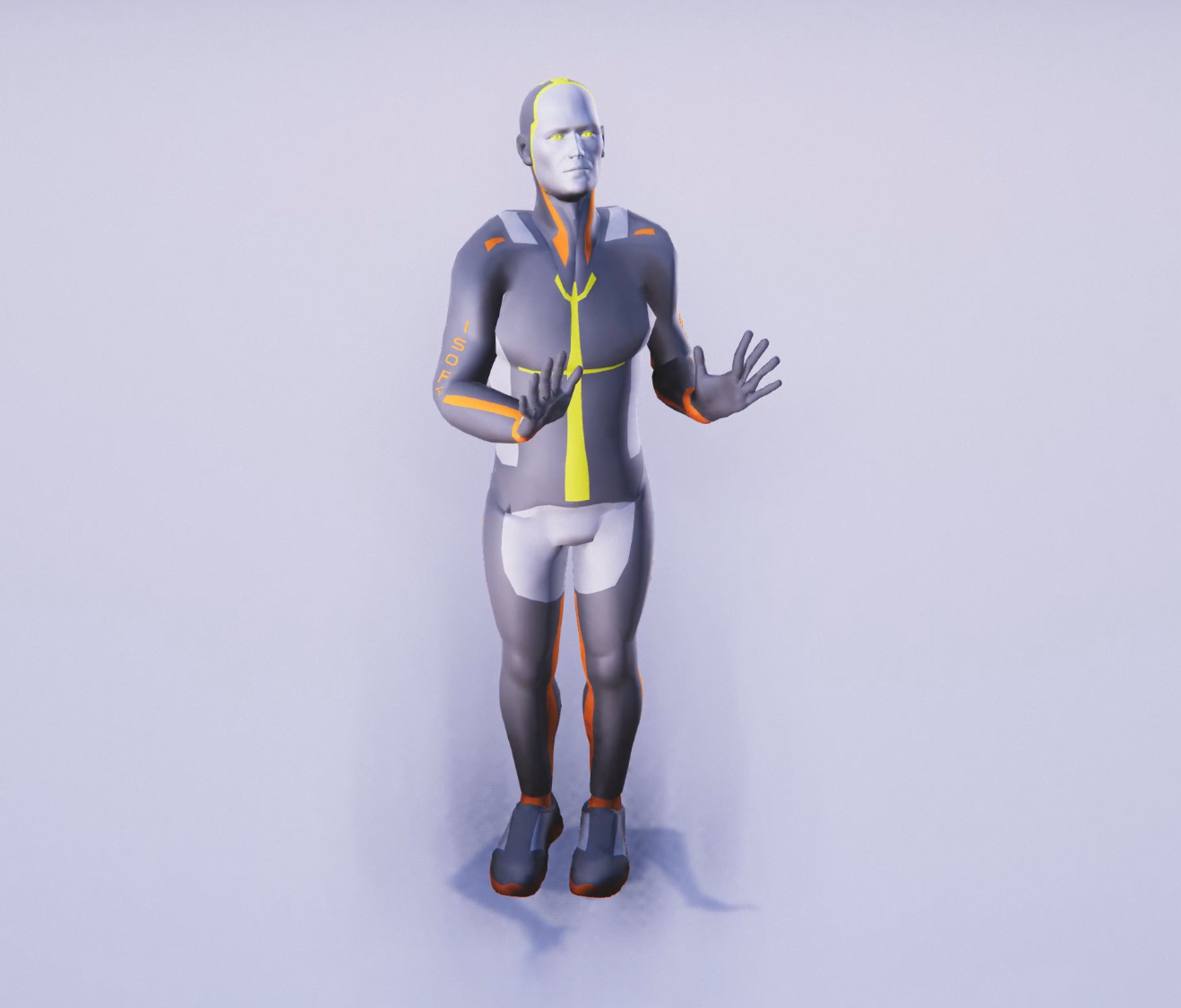}}
		
		{\tiny Click the image to watch the video}
	\end{center}
}

Illustration of gesture generation with synthesized speech from Microsoft Azure introducing the topic.

\subsection{Illustration of Character Motion}

\begin{center}
	{
		\centering
		\href{https://www.youtube.com/watch?v=9IIIZP3EJLg}{
			\includegraphics[width=\linewidth]{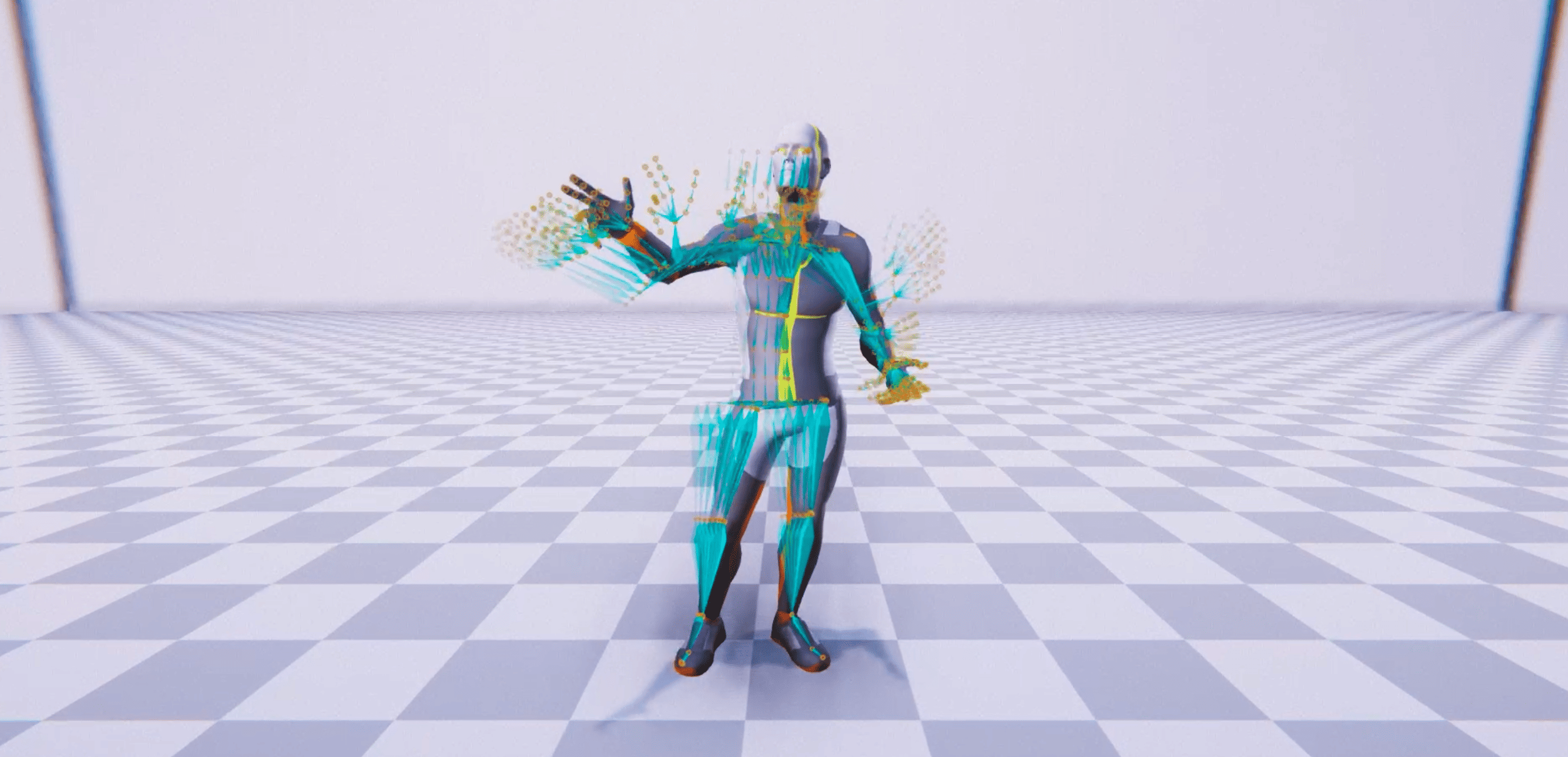}}
		
		{\tiny Click the image to watch the video}
	}
\end{center}
Illustration of gestures extracted from 6 frames before and 6 frames after the target gesture.